\documentclass[twocolumn]{aastex63}

\usepackage{textcomp}
\usepackage{multirow}
\usepackage{bm}
\usepackage[caption=false]{subfig}

\shorttitle{NIR extinction curves}
\shortauthors{Decleir et al.}

\graphicspath{{./}{figures/}}

\begin{document}

\title{SpeX near-infrared spectroscopic extinction curves in the Milky Way}

\correspondingauthor{Marjorie Decleir}
\email{mdecleir@stsci.edu}

\author[0000-0001-9462-5543]{Marjorie Decleir}
\affiliation{Space Telescope Science Institute, 3700 San Martin Drive, Baltimore, MD, 21218, USA}

\author[0000-0001-5340-6774]{Karl D. Gordon}
\altaffiliation{Visiting Astronomer at the Infrared Telescope Facility, which is operated by the University of Hawaii under contract 80HQTR19D0030 with the National Aeronautics and Space Administration.}
\affiliation{Space Telescope Science Institute, 3700 San Martin Drive, Baltimore, MD, 21218, USA}
\affiliation{Sterrenkundig Observatorium, Universiteit Gent,
  Gent, Belgium}

\author[0000-0003-0123-0062]{Jennifer E. Andrews}
\altaffiliation{Visiting Astronomer at the Infrared Telescope Facility, which is operated by the University of Hawaii under contract 80HQTR19D0030 with the National Aeronautics and Space Administration.}
\affiliation{Gemini Observatory/NSF’s NOIRLab, 670 N. A’ohoku Place, Hilo, HI 96720, USA}

\author[0000-0002-0141-7436]{Geoffrey~C.~Clayton}
\altaffiliation{Visiting Astronomer at the Infrared Telescope Facility, which is operated by the University of Hawaii under contract 80HQTR19D0030 with the National Aeronautics and Space Administration.}
\affiliation{Department of Physics and Astronomy, Louisiana State University, Baton Rouge, LA, 70803 USA}

\author[0000-0001-7780-3352]{Michael C. Cushing}
\affiliation{Ritter Astrophysical Research Center, Department of Physics and Astronomy, University of Toledo, 2801 W. Bancroft St., Toledo, OH 43606, USA}

\author{Karl A.\ Misselt}
\altaffiliation{Visiting Astronomer at the Infrared Telescope Facility, which is operated by the University of Hawaii under contract 80HQTR19D0030 with the National Aeronautics and Space Administration.}
\affiliation{Steward Observatory, University of Arizona, 933 North Cherry Avenue, Tucson, AZ 85721, USA}

\author[0000-0001-8102-2903]{Yvonne~Pendleton}
\affiliation{NASA Ames Research Center, Moffett Field, CA 94035, USA}

\author{John Rayner}
\altaffiliation{Visiting Astronomer at the Infrared Telescope Facility, which is operated by the University of Hawaii under contract 80HQTR19D0030 with the National Aeronautics and Space Administration.}
\affiliation{Institute for Astronomy, University of Hawaii, 2680 Woodlawn Drive, Honolulu, HI, USA 96822}

\author[0000-0002-9123-0068]{William D. Vacca}
\altaffiliation{Visiting Astronomer at the Infrared Telescope Facility, which is operated by the University of Hawaii under contract 80HQTR19D0030 with the National Aeronautics and Space Administration.}
\affiliation{SOFIA-USRA, NASA Ames Research Center, MS N232-12, Moffett Field, CA , USA}

\author[0000-0001-8539-3891]{D.~C.~B.~Whittet}
\affiliation{Department of Physics, Applied Physics and Astronomy, 
  Rensselaer Polytechnic Institute, 110 Eighth Street, Troy, NY 12180, USA}

\begin{abstract}

Interstellar dust extinction curves provide valuable information about the dust properties, including the composition and size of the dust grains\replaced{. In addition, extinction curves}{, and} are essential to correct observations for the effects of interstellar dust.\deleted{ They are found to vary between different sightlines in the Milky Way.} In this work, we measure a representative sample of near-infrared (NIR, 0.8--5.5\,\micron) spectroscopic extinction curves \added{for the first time}, enabling us to investigate the extinction at wavelengths where it is usually only measured in broad photometric bands. We use IRTF/SpeX spectra of a sample of \deleted{25 }reddened and \deleted{15 }comparison stars to measure \added{15} extinction curves with the pair method. Our sample spans $A(V)$ values from 0.78 to 5.65 and $R(V)$ values from \replaced{2.33}{2.43} to 5.33. We confirm that the NIR extinction curves are well fit by a power law, with indices and amplitudes differing from sightline to sightline. \replaced{The}{Our} average diffuse NIR extinction curve can be represented by a single power law with index $\alpha=1.7$,\deleted{ similar to previous works,} \added{but because of the sightline-to-sightline variations, the shape of any average curve will depend on the parental sample.} We find that most of the variation in our sample can be linked to the \added{ratio of} total-to-\replaced{selection}{selective} extinction $R(V)$, a \added{rough} measurement of the \added{average} dust grain size\deleted{s}. Two sightlines in our sample clearly show the ice extinction feature at 3\,\micron, which can be \replaced{approximated}{fitted} by a modified Drude profile. We find tentative ice detections with slightly over $3\sigma$ significance in two other sightlines. In \replaced{the}{our} average diffuse extinction curve, we \replaced{did not detect this feature with}{measure} a $3\sigma$ upper limit of \replaced{0.0021\,$A(3\,\micron)/A(V)$}{$A(\text{ice})/A(V)=0.0021$} \added{for this ice feature}.

\end{abstract}


\section{Introduction} \label{sec:intro}

Interstellar dust plays a significant role in several physical and chemical processes in the interstellar medium (ISM). The dust grains regulate the temperature of the gas, catalyze the formation of molecular hydrogen, and provide a reservoir of heavy elements. Furthermore, dust absorbs and scatters \deleted{a large amount of the} starlight at ultraviolet (UV), optical, \deleted{ and} near- \added{and mid-}infrared (NIR\added{--MIR}) wavelengths, and re-emits this energy at infrared (IR) wavelengths. This reprocessing of electromagnetic radiation highly affects our ability to study the universe at these wavelengths, and alters the observed spectral energy distribution (SED) of celestial objects. It is therefore of utmost importance to understand the interplay between dust and starlight.

The extinction of starlight by dust, i.e. the combined effect of absorption and scattering, is wavelength dependent and can be described by an \textit{extinction curve}. Extinction curves are not only required to correct the observations of a range of astrophysical objects for the effects of dust, but also provide insights into the properties of the interstellar dust: the continuum shape contains details about the size distribution of the grains, while extinction features reveal the composition of the dust grains.
\added{Over the past decades, several techniques have been developed to measure extinction curves, which can broadly be categorized in two types: those using individual sightlines, and those using an ensemble of stars. Both types are complimentary and come with their own advantages and disadvantages.}

\replaced{Extinction curves are commonly measured using the}{A commonly used method of the first type is} the \textit{pair method} \citep{1965ApJ...142.1683S, 1983ApJ...266..662M}: the observed SED of a reddened star is compared to (i.e. divided by) the observed SED of an unreddened comparison star with similar stellar properties. The difference in their SEDs is then attributed to the dust in the line-of-sight towards the reddened star.
\added{This method has been used extensively to measure extinction curves in samples of diffuse and dense sightlines in the Milky Way \citep[e.g.,][]{1985ApJ...288..618R, 1986ApJ...307..286F, 1988ApJ...328..734F, 1989ApJ...345..245C, 1990ApJ...357..113M, 2003ApJ...592..947C, 2004ApJ...616..912V, 2009ApJ...705.1320G,2021ApJ...916...33G}, as well as in the Magellanic Clouds \citep[e.g.,][]{2003ApJ...594..279G}. The main benefit of using comparison stars is that it does not require any modeling or absolute flux calibrations. One caveat is that the extinction measurements are relative (to the comparison star), and often only a limited number of comparison stars are available which makes it more difficult to exactly match the stellar properties of the reddened star.
It is also possible to use stellar atmosphere models instead of observed comparison stars with which to compare the SED (or the colors) of the reddened star. This has been done in the Milky Way \citep[by e.g.,][]{,2005AJ....130.1127F,2007ApJ...663..320F,2009ApJ...699.1209F,2019ApJ...886..108F,2020ApJ...891...67M}, in M31 \citep{2015ApJ...815...14C}, and in the Small Magellanic Cloud \citep{2012A&A...541A..54M}. This method makes it easier to closely match the properties of the reddened star, but its accuracy relies on a precise absolute flux calibration and accurate stellar atmosphere models.}

\added{The second technique to measure extinction uses an ensemble of stars, rather than individual sightlines. \cite{2006ApJ...638..839N} and \cite{2017ApJ...849L..13A}, for example, used the positions of red clump (RC) stars in color-magnitude diagrams (CMD) as a tracer of the extinction and reddening toward the Galactic center (GC). This technique is referred to as the \textit{RC method}, and a variant was used by \cite{2009ApJ...696.1407N} to determine the extinction curve toward the GC using magnitudes of bulge RC stars and colors of red giant branch (RGB) stars. \cite{2018AA...610A..83N, 2019AA...630L...3N} explored different methods based on RC stars to measure the extinction toward the GC, including using a combination of stellar atmosphere models and an extinction grid, using a fixed extinction, using a color-color diagram (CCD), and using a CMD. \cite{2005ApJ...619..931I}, \cite{2009MNRAS.400..731S} and \cite{2020MNRAS.496.4951M} also used the position of RC stars in a CCD to measure the extinction in the Galactic plane. 
Finally, \cite{2011ApJ...737...73F} derived the extinction curve toward the GC using nebular hydrogen emission lines.

The ensemble methods listed above are very valuable to measure extinction toward the GC in the NIR, especially given the large number of stars and strong extinction effects. However, it is not possible to measure UV and optical extinction, because the GC cannot be detected at those wavelengths due to its very high extinction. To obtain a detailed understanding of the interstellar dust properties (e.g. size, composition), it is critical to combine multi-wavelength extinction measurements from UV to MIR. Furthermore, dust depletions, which are measured from UV absorption line spectra of bright nearby stars, provide unique insights into the dust properties. Since UV--optical extinction measurements and depletions can only be measured in the local ISM, we focus on the local ISM (i.e. within 3\,kpc) in this paper.}

Studies of the Milky Way extinction curve at UV and optical wavelengths\deleted{ are omnipresent and} have shown that the extinction curve (and thus the dust properties) is not universal, but varies from sightline to sightline \cite[e.g.,][]{1989ApJ...345..245C,1992ApJ...398..610M,2004ApJ...616..912V,2009ApJ...705.1320G,2019ApJ...886..108F,2020ApJ...891...67M}.\deleted{ However, there is a paucity of studies at longer wavelengths.} \replaced{For a very long time}{However, for many years}, it was believed that, while extinction curves vary significantly at UV and optical wavelengths in different regions of the Galaxy, they are rather uniform at \replaced{longer}{NIR} wavelengths \deleted{$>$0.7\,\micron} \citep[e.g.,][]{1980MNRAS.192..359J, 1985ApJ...288..618R, 1987MNRAS.227..943S, 1989ESASP.290...93D, 1989ApJ...345..245C, 1990ApJ...357..113M}. Several studies found that the IR extinction curve from \replaced{$\sim0.7$}{$\sim0.9$} to $\sim5$\,\micron\ can be closely represented by a power law ($A(\lambda)\propto\lambda^{-\alpha}$), with reported values for the index $\alpha\simeq1.75$ \citep{1989ESASP.290...93D}, $\alpha\simeq1.6$ \citep{1989ApJ...345..245C}, and $\alpha\simeq1.8$ \citep{1990ApJ...357..113M} \added{in the local ISM}.
\added{More recent extinction measurements toward the GC \added{and in the Galactic disk} showed steeper NIR extinction curves (between $\sim1.2$ and $\sim2.2$\,\micron) with indices $\alpha\simeq1.99$ \citep{2006ApJ...638..839N}, $\alpha\simeq2$ \citep{2009ApJ...696.1407N}, $\alpha\simeq2.14$ \citep{2009MNRAS.400..731S}, $\alpha\simeq2.11$ \citep{2011ApJ...737...73F}, $\alpha\simeq2.47$ \citep{2017ApJ...849L..13A}, $\alpha\simeq2.30$ \citep{2018AA...610A..83N}, $\alpha\simeq2.32$ \citep{2019AA...630L...3N}, and $\alpha\simeq2.27$ \citep{2020MNRAS.496.4951M}.
Some of these \deleted{GC} studies reported a dependence of the power law index on the wavelength region. More specifically, \cite{2005ApJ...619..931I} and \cite{2009ApJ...696.1407N} found a flattening of the extinction curve at wavelengths beyond $\sim3$\,\micron, \cite{2011ApJ...737...73F} obtained a flatter extinction curve beyond $\sim3.7$\,\micron\, and \cite{2019AA...630L...3N} found different power law indices between JH (1.27--1.65\,\micron, $\alpha\simeq2.43$) and HK\textsubscript{S} (1.65--2.16\,\micron, $\alpha\simeq2.23$).
\cite{2006ApJ...638..839N, 2009ApJ...696.1407N}, \cite{2011ApJ...737...73F}, and \cite{2017ApJ...849L..13A} found that the NIR extinction curve changes from one sightline to another, in contrast to what was previously believed. On the other hand, \cite{2005ApJ...619..931I}, \cite{2009MNRAS.400..731S}, \cite{2018AA...610A..83N, 2019AA...630L...3N} and \cite{2020MNRAS.496.4951M} saw no significant variation in the NIR extinction curve as a function of sightline in the Galactic regions they studied.}

\deleted{However,} \replaced{all}{All} of the \added{above-mentioned} IR studies \added{(except for \cite{2011ApJ...737...73F})} are based on a limited number of broad-band photometric data points (usually a subset of the IJHKLM bands), in contrast to most of the UV\deleted{/optical} studies \replaced{mentioned}{listed} above \added{and the recent optical measurements of \cite{2019ApJ...886..108F} and \cite{2020ApJ...891...67M}} which use spectra.\deleted{ Furthermore, they report quite large uncertainties in the extinction measurements above 3\,\micron.}
\added{As, for example, suggested by \cite{2020MNRAS.496.4951M}, there is a limit to what can be studied about the extinction curve with photometry.} Recently, \cite{2021ApJ...916...33G} characterized the \deleted{mid-infrared (}MIR\deleted{)} extinction curve based on Spitzer photometry (3.6\textendash24\,\micron) and spectra (5\textendash37\,\micron). They showed that the average \added{diffuse} Milky Way extinction curve at these wavelengths can be represented by a power law with index $\alpha=1.48$ (and two modified Drude profiles for the silicate features). However, they also found large variations in the shape of the extinction curve between different sightlines, proving that the IR extinction is not uniform within the Galaxy.

With the work presented in this paper, we fill the gap between the spectral UV/optical extinction studies and the recent MIR results from \cite{2021ApJ...916...33G}, by measuring extinction curves at wavelengths between 0.8 and 5.5\,\micron, for the first time using NIR spectra (instead of broadband data). We utilize SpeX spectra for a sample of 25 reddened stars and 15 comparison stars, and measure the extinction towards the reddened stars with the pair method. With this data set, we are not only able to characterize the shape of an average \added{diffuse} Milky Way NIR extinction curve at spectroscopic resolution, but also to study the variations between the different sightlines in our sample.

Section \ref{sec:data} describes the sample, and the processing of the SpeX spectra. In Section \ref{sec:measure} we explain how we measured extinction curves from these data. The fitting of the extinction curves is outlined in Section \ref{sec:fitting}. Section \ref{sec:results} presents and discusses the results of the fitting, the correlation between the fitting parameters, the average diffuse Milky Way extinction curve with a comparison to other studies and dust grain models, the correlation of the sightline variations with $R(V)$, and the observed extinction features. Finally, Section \ref{sec:conclusions} summarizes this work.

\section{Data processing} \label{sec:data}

\subsection{Sample}
\label{sec:sample}
Our sample consists of 15 comparison and 25 reddened Milky Way OB stars. They are listed in Table~\ref{tab:sample} with their spectral type, \added{B and V-band photometry, and distance, all obtained from the literature (see references in the table).} OB stars are particularly suited for determining extinction curves, because their spectra exhibit fewer stellar lines compared to later type stars. Furthermore, these stars are luminous at UV--MIR wavelengths, which are the wavelengths of interest to study dust extinction. \added{The comparison stars were selected to have very little dust along their line-of-sight, as can be seen from their estimated reddening, $E(B-V)$, listed in Table~\ref{tab:sample}. To obtain these reddening values, we calculated the observed ($B-V$)-color for every comparison star from its B and V-band photometry (Table~\ref{tab:sample}), and subtracted the intrinsic ($B-V$)-color from Table~1 in \cite{1970A&A.....4..234F} for the corresponding spectral type and luminosity class (also given in our Table~\ref{tab:sample}). Note that these values are not used in our analysis, but just given as a reference.\footnote{For star \object{HD164794} (spectral type O4), no intrinsic ($B-V$)-color was available, but we did not use this star to measure extinction curves.}} The sample of reddened sightlines was chosen so that it represents a large range in $R(V)$ values (\replaced{2.3}{2.4}--5.3), which probe the dust grain size along the line-of-sight. The $E(B-V)$ values range from 0.2 to 1.6, and $A(V)$ values from 0.8 to 5.6.\footnote{$R(V)$, $E(B-V)$ and $A(V)$ will be defined in Section \ref{sec:measure}.}

\begin{table*}
\centering
\caption{Sample stars with their spectral type\replaced{ and its reference.}{, B and V-band photometry, distance, and corresponding references. For the comparison stars, also an estimate of their reddening, $E(B-V)$, is given.}}\label{tab:sample}
\begin{tabular}{llccccccc}
\hline\hline
\textbf{star} & \textbf{type}            & \textbf{type} & \textbf{B} & \textbf{V} & \textbf{BV} & \textbf{dist.} & \textbf{dist.} & $\bm{E(B-V)}$\\
 &           & \textbf{ref.} & \textbf{[mag]} & \textbf{[mag]} & \textbf{ref.} & \textbf{[pc]} & \textbf{ref.} & \\
\hline
\multicolumn{9}{c}{\textbf{comparison stars}}                 \\
\hline
\object{HD003360}*      & B2IV        & {[}1{]}   & $3.47\pm0.03$ & $3.66\pm0.03$ & [12] & 109 & [25] & 0.05\\
\object{HD031726}*      & B1V         & {[}1{]}   & $5.941\pm0.002$ & $6.147\pm0.001$ & [13] & 389 & [25] & 0.05   \\
\object{HD032630}      & B3V         & {[}1{]}   & $3.00\pm0.03$ & $3.18\pm0.02$ & [12] & 63 & [25] & 0.02 \\
\object{HD034759}*      & B5V         & {[}1{]}   & $5.08\pm0.02$ & $5.22\pm0.02$ & [14] & 179 & [25] &0.02\\
\object{HD034816}*      & B0.5IV      & {[}1{]}   & $4.04\pm0.03$ & $4.29\pm0.02$ & [12] & 270 & [25] & 0.03\\
\object{HD036512}      & B0V         & {[}1{]}   & $4.36\pm0.03$ & $4.62\pm0.02$ & [12] & \nodata & \nodata & 0.04\\
\object{HD042560}*      & B3IV        & {[}1{]}   & $4.31\pm0.03$ & $4.48\pm0.03$ & [12] & 210 & [25] & 0.03\\
\object{HD047839}      & O7V((f))z var   & {[}2{]}   & $4.42\pm0.02$ & $4.66\pm0.02$ & [12] & 609 & [26] & 0.08\\
\object{HD051283}*      & B3II-III    & {[}3{]}   & $5.15\pm0.03$ & $5.32\pm0.02$ & [15] & 736 & [25] & 0.01 \\
\object{HD078316}      & B8IIIp      & {[}4{]}   & $5.13\pm0.02$ & $5.24\pm0.02$ & [16] & 154 & [25] & -0.01 \\
\object{HD091316}      & B1Iab       & {[}1{]}   & $3.71\pm0.02$ & $3.85\pm0.02$ & [12] & 334 & [26] & 0.05 \\
\object{HD164794}      & O4V((f))z   & {[}2{]}   & $5.97\pm0.03$ & $5.97\pm0.02$ & [12] & 950 & [26] & \nodata \\
\object{HD188209}*      & O9.5Ia      & {[}1{]}   & $5.55\pm0.03$ & $5.62\pm0.02$ & [12] & 1112 & [26] & 0.20 \\
\object{HD204172}      & B0Ib        & {[}1{]}   & $5.85\pm0.02$ & $5.93\pm0.02$ & [17] & \nodata & \nodata & 0.16  \\
\object{HD214680}*      & O9V         & {[}1{]}   & $4.68\pm0.03$ & $4.88\pm0.02$ & [12] & 359 & [25] & 0.11  \\
\hline
\multicolumn{9}{c}{\textbf{reddened stars}}                   \\
\hline
\object[BD +56 524]{BD+56d524}*     & B1V         & {[}5{]}   & $10.09\pm0.01$ & $9.75\pm0.01$ & [11] & 2349 & [25] \\
\object{HD013338}*      & B1V         & {[}3{]}   & $9.27\pm0.01$ & $9.03\pm0.01$ & [11] & 2425 & [25] \\
\object{HD014250}      & B0.5V:n     & {[}3{]}   & $9.28\pm0.01$ & $8.96\pm0.01$ & [11]  & 1245 & [25] \\
\object{HD014422}      & B1V:pe      & {[}3{]}   & $9.53\pm0.01$ & $9.03\pm0.01$ & [18]  & \nodata & \nodata\\
\object{HD014956}*      & B2Ia        & {[}3{]}   & $7.91\pm0.03$ & $7.19\pm0.02$ & [19] & \nodata  & \nodata\\
\object{HD017505}*      & O6.5III((f))n+O8V       & {[}6{]} & $7.46\pm0.01$ & $7.06\pm0.01$ & [11]& \nodata & \nodata\\
\object{HD029309}*      & B2V         & {[}7{]}   & $7.42\pm0.01$ & $7.10\pm0.01$ & [20] & 544 & [25] \\
\object{HD029647}*      & B7IV        & {[}8{]}   & $9.22\pm0.03$ & $8.31\pm0.02$ & [21] & 155 & [25] \\
\object{HD034921}      & B0IVpe      & {[}3{]}   & $7.65\pm0.01$ & $7.51\pm0.01$ & [22]  & 1327 & [25] \\
\object{HD037020}      & O8Vn        & {[}7{]}   & $6.72\pm0.01$ & $6.72\pm0.01$ & [11]  & 421 & [25]\\
\object{HD037022}      & O7Vp        & {[}6{]}   & $5.13\pm0.02$ & $5.13\pm0.01$ & [11]  & 430 & [26]\\
\object{HD037023}      & B0.5Vp      & {[}5{]}   & $6.77\pm0.01$ & $6.69\pm0.01$ & [11]  & 472 & [25]\\
\object{HD037061}*      & B1V         & {[}9{]}   & $7.09\pm0.01$ & $6.83\pm0.01$ & [11] & 523 & [25]\\
\object{HD038087}*     & B3II        & {[}10{]}  & $8.42\pm0.01$ & $8.30\pm0.01$ & [11]  & 339 & [25]\\
\object{HD052721}      & B2Vne       & {[}7{]}   & $6.64\pm0.06$ & $6.58\pm0.05$ & [11]  & \nodata & \nodata\\
\object{HD156247}*      & B3V         & {[}10{]}  & $5.96\pm0.04$ & $5.91\pm0.04$ & [11]  & 267 & [25] \\
\object{HD166734}      & O7.5Iabf    & {[}6{]}   & $9.51\pm0.01$ & $8.42\pm0.01$ & [22]  & 3418 & [26]\\
\object{HD183143}*      & B7Ia        & {[}3{]}   & $8.08\pm0.01$ & $6.86\pm0.01$ & [23] & \nodata & \nodata \\
\object{HD185418}*      & B0.5V       & {[}3{]}   & $7.67\pm0.01$ & $7.45\pm0.01$ & [11] & 755 & [25] \\
\object{HD192660}*      & B0Ia        & {[}3{]}   & $8.05\pm0.04$ & $7.38\pm0.03$ & [15] & 2003 & [25]\\
\object{HD204827}*      & B0V         & {[}3{]}   & $8.76\pm0.01$ & $7.95\pm0.01$ & [22] & 1646 & [26] \\
\object{HD206773}      & B0V:pe      & {[}3{]}   & $7.01\pm0.02$ & $6.79\pm0.02$ & [24]  & 958 & [25]\\
\object{HD229238}*      & B0Iab       & {[}3{]}   & $9.78\pm0.01$ & $8.88\pm0.01$ & [22] & 1473 & [25] \\
\object{HD283809}*      & B3V         & {[}8{]}   & $12.14\pm0.03$ & $10.72\pm0.02$ & [21] & 326 & [25]\\
\object{HD294264}      & B3V         & {[}11{]}  & $9.83\pm0.01$ & $9.47\pm0.01$ & [11]  & 450 & [25]\\
\hline
\end{tabular}
\tablerefs{[1] \citet{1968ApJS...17..371L}; [2] \citet{2014ApJS..211...10S}; [3] \citet{1955ApJS....2...41M}; [4] \citet{1975A&AS...19...91L}; [5] \citet{1960BAN....15..255B}; [6] \citet{2011ApJS..193...24S}; [7] \citet{1968PASP...80..197G}; [8] \citet{2000ApJS..128..603M}; [9] \citet{1952ApJ...116..251S}; [10] \citet{1999MSS...C05....0H}; [11] \citet{2004ApJ...616..912V}; [12] \citet{1966CoLPL...4...99J}; [13] \citet{1990SAAOC..14...33M}; [14] \citet{1971AJ.....76.1058C}; [15] \citet{1983ApJS...52....7F}; [16] \citet{1962ApJ...136...35A}; [17] \citet{1977AJ.....82..431L}; [18] \citet{1955ApJ...122..429J}; [19] \citet{1967BOTT....4..149M}; [20] \citet{1974PASP...86..795G}; [21] \citet{1980SvAL....6..397S}; [22] \citet{1956ApJS....2..389H}; [23] \citet{1965ApJ...141..923J}; [24] \citet{1976PASP...88..865G}; [25] \citet{2018yCat.1345....0G}; [26] \citet{2009A&A...507..833M}.}
\tablecomments{\added{Stars with an * were used to measure extinction curves, as explained in Section 3.}}
\end{table*}

\subsection{SpeX NIR spectra}
\label{sec:spectra}
The NIR spectra have been obtained during an observational campaign at the 3.2\,m NASA Infrared Telescope Facility (IRTF) on Mauna Kea. Observations were obtained with SpeX, a medium-resolution 0.8\sbond5.5\,\micron\ spectrograph, in two modes: the Short wavelengths Cross-Dispersed mode (SXD, $\sim$0.8\sbond2.4\,\micron), and the Long wavelengths Cross-Dispersed mode (LXD, $\sim$1.9\sbond5.5\,\micron) \citep{2003PASP..115..362R}. The observations have been performed over several nights in 2005, 2006 and 2007. In addition, we used 7 spectra from the IRTF spectral library, observed over multiple years, starting in 2000 and ending in 2003.

\begin{figure*}[t]
\plotone{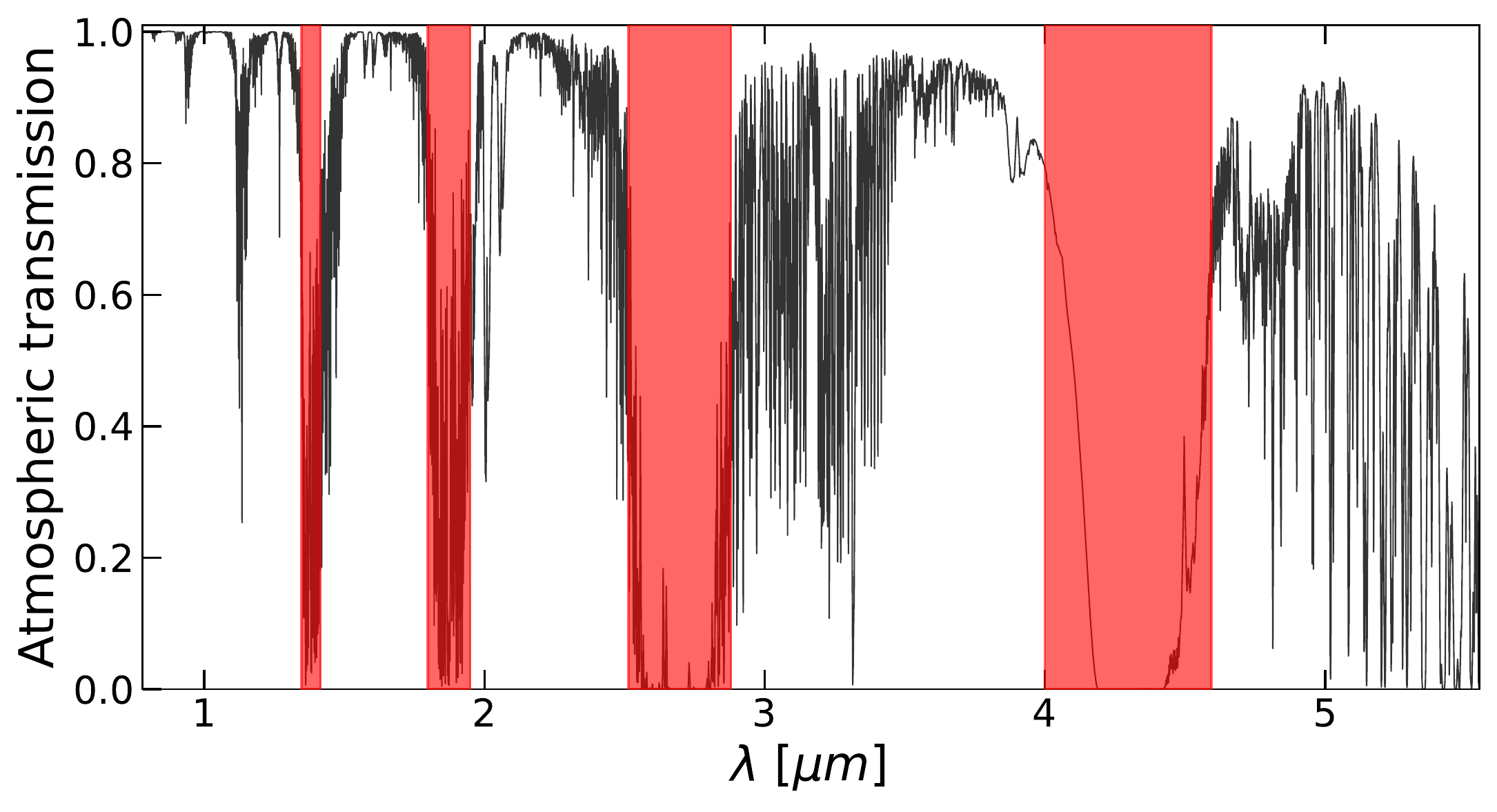}
\caption{An atmospheric transmission model, obtained from the Spextool database, and computed by \cite{2004PASP..116..362C} with the atmospheric transmission tool ATRAN \citep{1992nstc.rept.....L}. The red shading indicates wavelength regions where the atmospheric transmission is very low, and the spectra are significantly affected by the telluric absorption. These regions were masked in the spectra. \label{fig:atmos}}
\end{figure*}

We reduced and processed the spectra with Spextool (SPectral EXtraction TOOL) (v.~4.1), an IDL-based data reduction package written by \cite{2004PASP..116..362C}. The data reduction started with extracting the spectra from the raw data, and the flat fielding and wavelength calibration of these spectra, using the \texttt{xspextool} GUI. Subsequently, multiple spectra of the same star were combined using the \texttt{xcombspec} GUI. Next, the stellar spectrum was corrected for the telluric absorption and the instrument throughput with the \texttt{xtellcor} program \citep{2003PASP..115..389V}. For this step, observations of a standard A0V star, and a high resolution model of Vega were used. After this, the different orders of the corrected spectrum were scaled and merged using the task \texttt{xmergeorders}, which results in one SXD and one LXD spectrum per star. It has to be noted that some wavelength regions of the spectra still suffered significantly from telluric absorption effects, despite the corrections applied with Spextool. These regions are shaded in red in Fig.~\ref{fig:atmos}, and were masked and excluded from all further analyses (such as measuring and fitting the extinction curves).

\begin{figure*}[ht]
\includegraphics[width=0.97\textwidth]{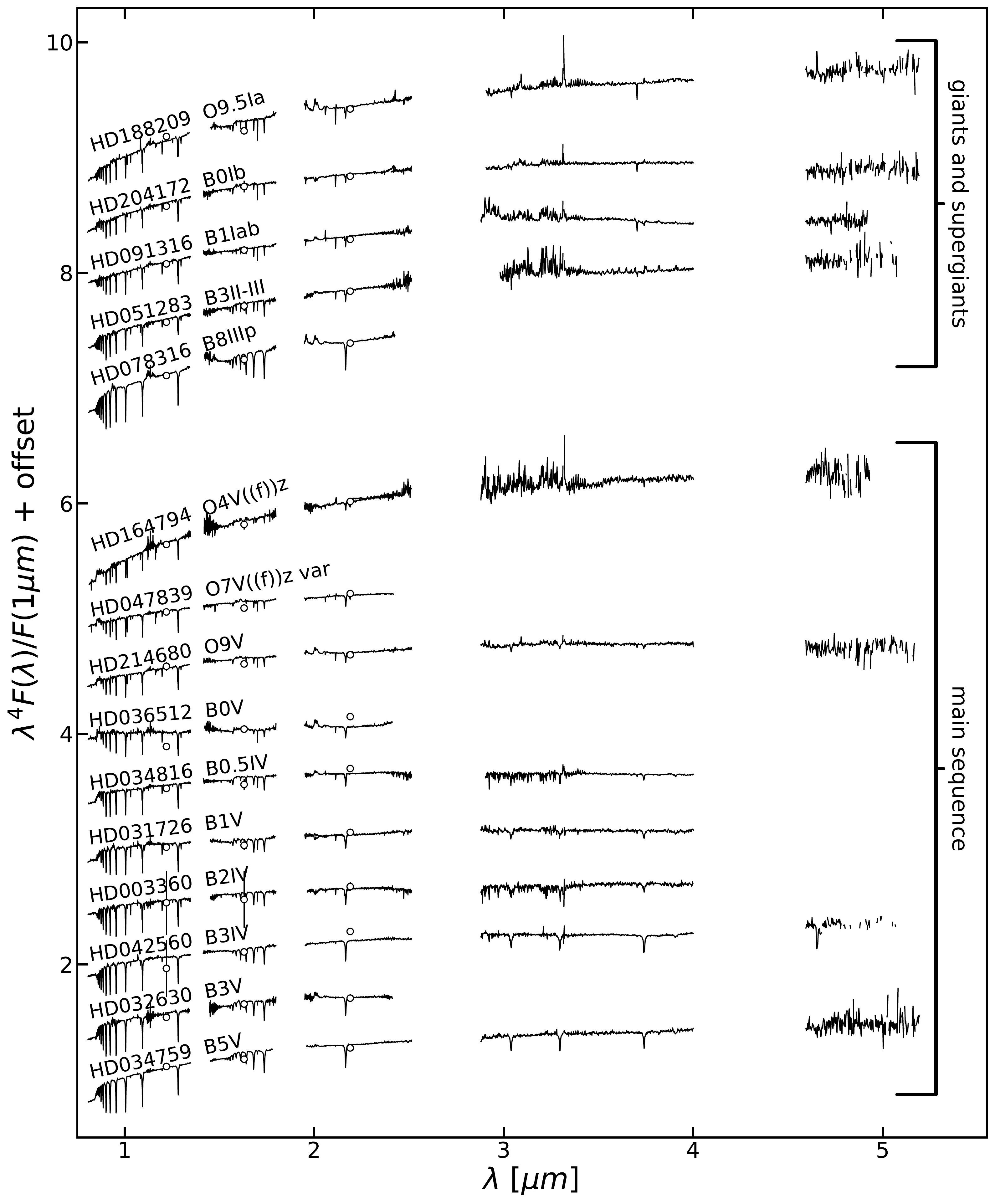}
\caption{The NIR spectra of the 15 comparison stars, ordered by luminosity class (giants and supergiants at the top, and main sequence stars at the bottom), and by spectral class (O4--B8 from top to bottom). The SpeX spectra are shown as solid lines and the photometric data (JHK\textsubscript{S}) are shown as open circles. All fluxes are normalized to the flux at 1\,\micron, and multiplied by $\lambda^4$ to remove the strongly decreasing Rayleigh-Jeans tail. \label{fig:comp_stars}}
\end{figure*}

\begin{figure*}[t]
\includegraphics[width=0.98\textwidth]{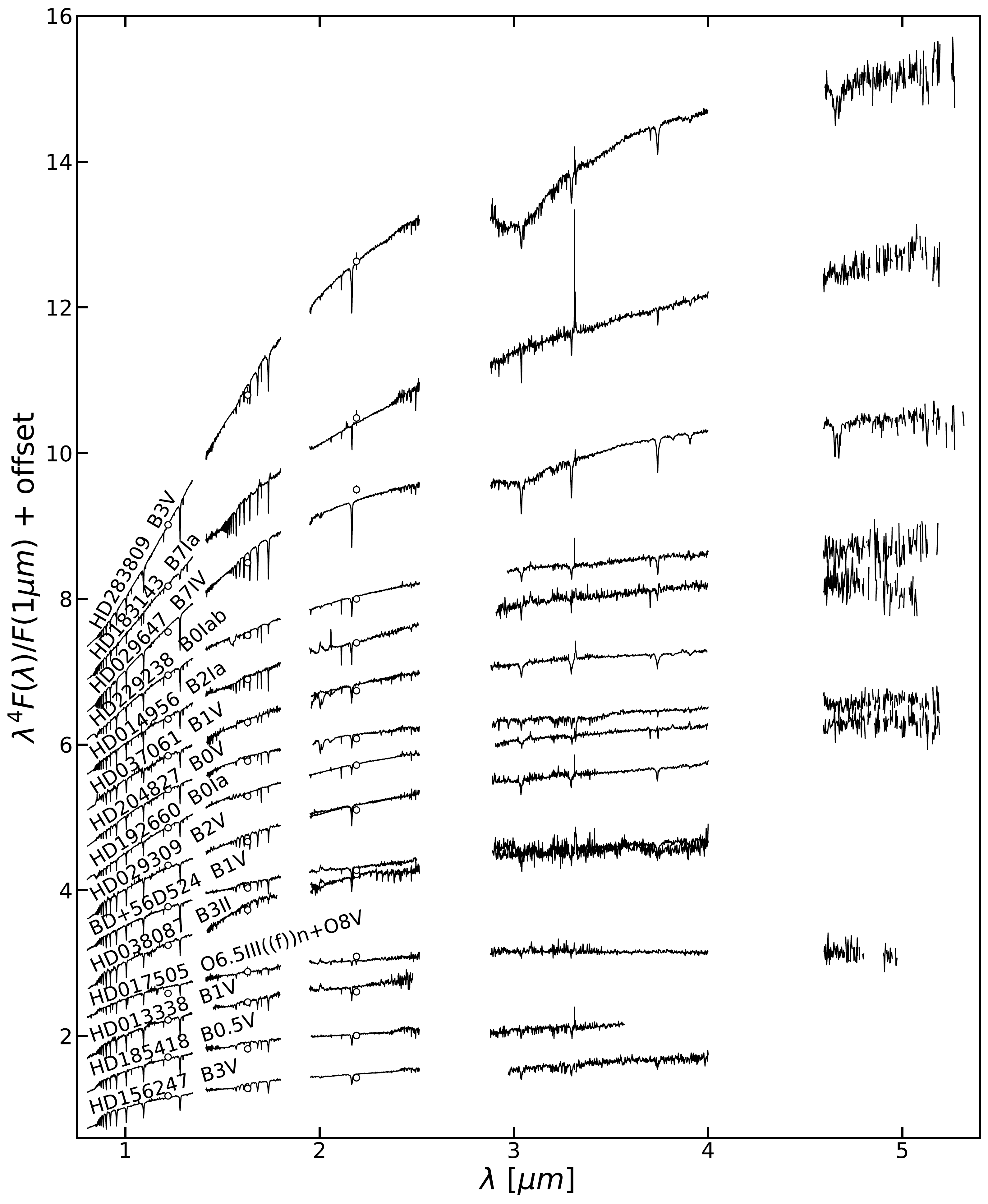}
\caption{The NIR spectra of the 15 reddened stars that were used to measure extinction curves, ordered by $A(V)$. The SpeX spectra are shown as solid lines and the photometric data (JHK\textsubscript{S}) are shown as open circles. All fluxes are normalized to the flux at 1\,\micron, and multiplied by $\lambda^4$ to remove the strongly decreasing Rayleigh-Jeans tail. \label{fig:red_stars}}
\end{figure*}

\begin{figure*}[t]
\includegraphics[width=\textwidth]{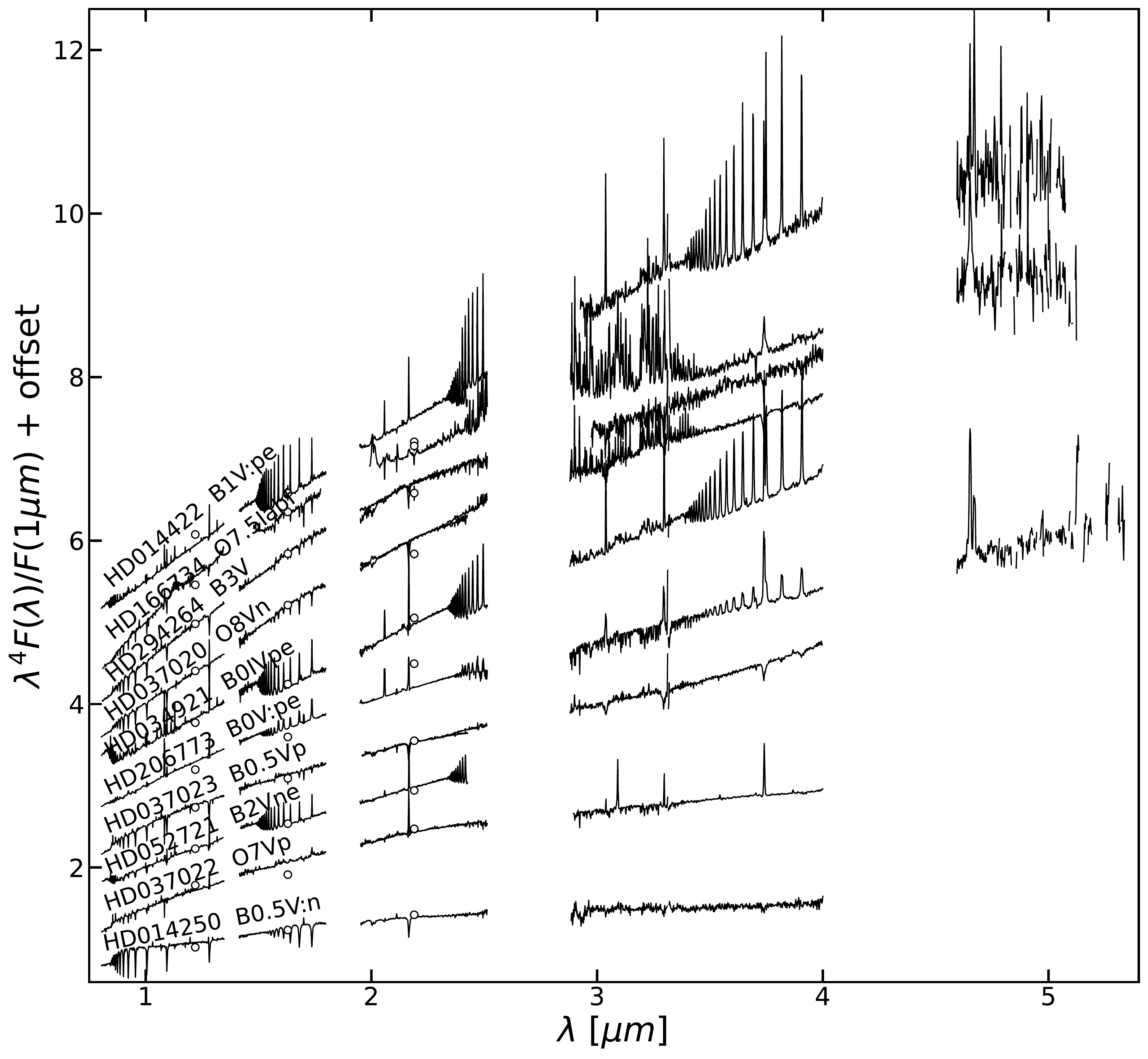}
\caption{The NIR spectra of the 10 reddened stars that could not be used to measure extinction curves because of emission lines or other peculiarities in the spectrum, ordered by steepness. The SpeX spectra are shown as solid lines and the photometric data (JHK\textsubscript{S}) are shown as open circles. All fluxes are normalized to the flux at 1\,\micron, and multiplied by $\lambda^4$ to remove the strongly decreasing Rayleigh-Jeans tail. \label{fig:bad_stars}}
\end{figure*}

\added{All spectra were then placed on the same wavelength grid (between 0.8 and 2.45\,\micron\ for SXD, and between 1.9 and 5.5\,\micron\ for LXD) with a resolution of 2000, using the \texttt{measure\_extinction} python package \citep{measureextinction}.} Finally, \added{with the same package}, the spectra were calibrated based on photometric data points\deleted{, using the \texttt{measure\_extinction} python package \citep{measureextinction}}. First, the SXD spectrum was scaled to match J, H, K\textsubscript{S} photometry from 2MASS \citep{2006AJ....131.1163S}, obtained from the \dataset[IRSA 2MASS All-Sky Point Source Catalog]{\doi{10.26131/IRSA2}} \citep{irsa2mass}. Subsequently, the LXD part of the spectrum was scaled to align with the SXD spectrum, resulting in one smooth NIR spectrum.\footnote{The LXD part could also be scaled using IRAC or WISE photometry, but this is not available for all stars in the sample. Furthermore, the uncertainties on the available photometric data points are fairly large, and using those data did not always result in a smooth NIR spectrum.} The final spectra of all stars \added{are electronically \dataset[available]{\doi{10.5281/zenodo.5802469}} \citep{decleir_marjorie_2022_5802469}, and} can be found in Figs.~\ref{fig:comp_stars}--\ref{fig:bad_stars}. They are plotted multiplied by $\lambda^4$ to remove the strongly decreasing Rayleigh-Jeans tail in this wavelength range. When plotted in this way, the spectrum of an unreddened (comparison) star flattens towards higher wavelengths.

The flux uncertainties that Spextool yields only include the photon noise and read noise. However, there are several other sources of uncertainty that need to be taken into account. For example, scaling and merging the different orders in the spectrum introduces some uncertainty, especially in those wavelength regions suffering from telluric absorption where it is difficult to measure the scaling factor based on the overlap of two adjacent orders. In addition, scaling the SXD spectrum based on the 2MASS photometry, and manually scaling the LXD spectrum to match the SXD spectrum, induces more uncertainty. Since it is very hard to quantify these uncertainties, we quadratically added a 1\% uncertainty to the photon and read noise to account for these uncertainties. 

Finally, we excluded data points with a signal-to-noise ratio (SNR) below 10 from all further analyses, limiting the wavelength range to $\sim$5.2\,\micron\ for most stars. \added{The median SNRs of the stellar spectra used in this work (indicated with an * in Table \ref{tab:sample}) are in the range 107--141 below 2.5\,\micron, 38--137 between 2.8 and 4\,\micron, and 18--72 beyond 4.5\,\micron, at a resolution of about 2000.}

\section{Measuring NIR extinction curves}
\label{sec:measure}
\begin{figure*}[ht]
\includegraphics[width=\textwidth]{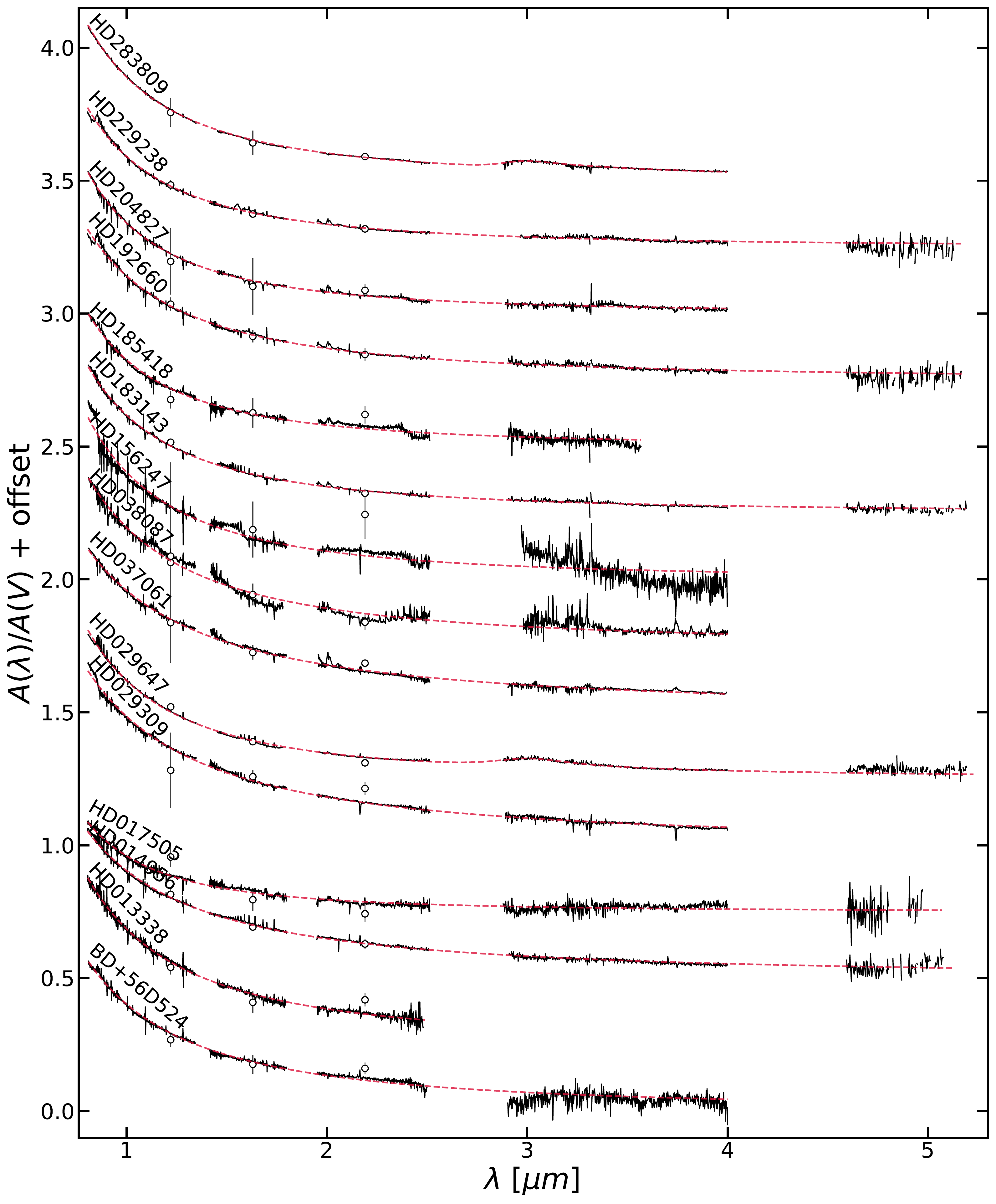}
\caption{The normalized NIR extinction curves for the 15 suitable sightlines in our sample. The SpeX spectra are shown as solid lines and the photometric data (JHK\textsubscript{S}) are shown as open circles. \added{The red dashed lines represent the fits to the data.} \label{fig:ext_curves}}
\end{figure*}

As mentioned in the introduction, we measured NIR extinction curves using the pair method: The spectrum of a reddened star was divided by the spectrum of an unreddened comparison star with similar stellar properties. \added{We want to point out that it is not possible to use stellar atmosphere models to measure NIR extinction curves with the pair method. The NLTE (non-local thermodynamic equilibrium) TLUSTY models \citep{2003ApJS..146..417L}, which are needed for OB stars, are not complete beyond 0.8\,\micron. Many stellar lines (including the upper Paschen lines) are missing \citep{2020ApJ...891...67M}, and also the continuum level is not reliable (priv.~comm.~I.~Hubeny).} \replaced{We matched}{Usually,} reddened and comparison stars \added{are matched} based on their spectral type, and if possible also on their luminosity class. \replaced{The spectral types and luminosity classes were taken from SIMBAD \citep{2000A&AS..143....9W}, and are reported in Table~\ref{tab:sample} together with their reference.}{The spectral types and luminosity classes listed in Table~\ref{tab:sample} were taken from SIMBAD \citep{2000A&AS..143....9W}.} It has to be noted\added{, however,} that for several stars, multiple spectral types are listed on SIMBAD (obtained from different references). Furthermore, spectral types might be different when derived using different methods or from different wavelength regions \citep[see e.g.,][]{1996ApJS..107..281H,1997AJ....114.1951S}. For these reasons, we \added{did not use the literature spectral types in Table~\ref{tab:sample} to match comparison and reddened stars, but merely list them as a reference.\footnote{Note that we used the literature spectral types to estimate the $E(B-V)$-values of the comparison stars, as explained in Section~\ref{sec:sample}.} Instead, we validated empirically which comparison star is the best match. We} tried \replaced{several}{all} comparison stars for every reddened star in our sample and retained the one that results in the smoothest extinction curve, i.e. that cancels out the stellar (hydrogen) lines \added{and jumps} as well as possible.  
Tables~\ref{tab:fit_results_diff} and \ref{tab:fit_results_dense} list the comparison star used to measure the extinction curve for every reddened star. Given the limited number of available comparison stars (only 15), the spectral match is not always perfect, and remaining stellar lines and jumps might be visible in some extinction curves (see Fig.~\ref{fig:ext_curves}).

The extinction curves were calculated with the \texttt{measure\_extinction} package \citep{measureextinction} using the following procedure. The \textit{absolute} dust extinction $A(\lambda)$ at a wavelength $\lambda$ is given by:
\begin{equation}
    A(\lambda) = -2.5 \: \text{log} \left[\frac{F(\lambda)_{\text{red}}}{F(\lambda)_{\text{comp}}} \right] + 5 \: \text{log} \left[ \frac{1/d_{\text{red}}}{1/d_{\text{comp}}} \right]
\end{equation}

\noindent where $F$ is the spectral flux density, $d$ is the distance to the star, ``red" refers to the reddened star, and ``comp" refers to the comparison star. Extinction is usually measured relative to a reference wavelength measurement in order to avoid the need of an accurate distance to the star. Most commonly, the V-band measurement is used as the reference. Thus, the \textit{relative} dust extinction $E(\lambda-V)$ (also called differential extinction, dust reddening or color excess) at wavelength $\lambda$ can be calculated as:
\begin{align}
E(\lambda-V) &= A(\lambda) - A(V)\label{eq:reddening} \\
&= -2.5 \: \text{log} \left[ \frac{F(\lambda)_{\text{red}}}{F(\lambda)_{\text{comp}}} \right] + 2.5 \: \text{log} \left[ \frac{F(V)_{\text{red}}}{F(V)_{\text{comp}}} \right]
\end{align}
\added{\noindent The V-band magnitudes of all stars are given in Table~\ref{tab:sample}.}

In order to compare between extinction curves from different sightlines, they must be normalized to the total level of extinction in that sightline, e.g. represented by $A(V)$, the absolute extinction in the V-band. The differential measurement $E(\lambda-V)$ can be converted into an absolute \textit{normalized} extinction measurement $A(\lambda)/A(V)$ from Eq.~\ref{eq:reddening}:
\begin{equation}
\label{eq:alav}
\frac{A(\lambda)}{A(V)} = \frac{E(\lambda-V)}{A(V)} + 1
\end{equation}

\noindent This conversion requires knowledge of $A(V)$, which we measured by fitting a power law to the NIR extinction curve, as discussed in the next section. \replaced{If}{Once} $A(V)$ is known, one can also compute the \textit{total-to-selective extinction} $R(V)$:
\begin{equation}
    R(V) = \frac{A(V)}{E(B-V)}
    \label{eq:RV}
\end{equation}
\noindent \added{with the color excess $E(B-V)$ calculated as
\begin{equation}
    E(B-V) = (m(B)_{\text{red}} - m(B)_{\text{comp}}) - (m(V)_{\text{red}}-m(V)_{\text{comp}})
\end{equation}

\noindent with $m(B)$ the apparent B-band magnitude and $m(V)$ the apparent V-band magnitude listed in Table~\ref{tab:sample}. The obtained values for $E(B-V)$ are given in Tables~\ref{tab:fit_results_diff} and \ref{tab:fit_results_dense}.\footnote{Several previous extinction studies pointed out that using monochromatic quantities to characterize an extinction curve is more appropriate than using the band-integrated equivalents we use here. However, as explained in App.~A of \cite{2018A&A...613A...9M}, and as visible in Fig.~3 of \cite{2013hsa7.conf..583M} and Table~4 of \cite{2019ApJ...886..108F}, these effects are very small for low-extinction OB stars, which we use in our sample.}} $R(V)$ probes the dust grain size along the line-of-sight, with larger values corresponding to sightlines dominated by larger dust grains \citep[e.g.,][]{1999PASP..111...63F}. On average, $R(V)=3.1$ in the Milky Way \added{diffuse ISM} \citep{1989ApJ...345..245C}.

\begin{figure}[ht]
\includegraphics[width=\columnwidth]{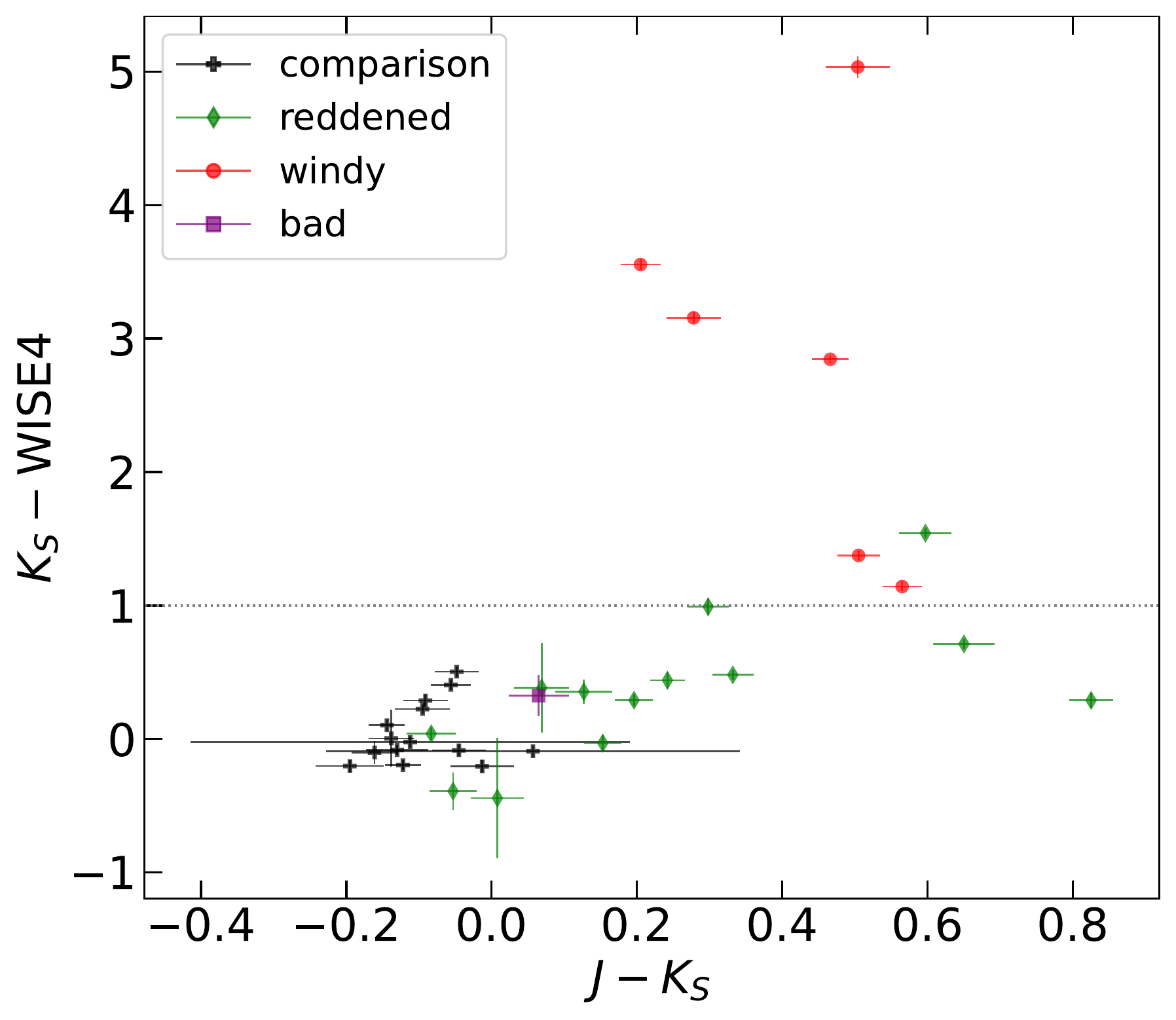}
\caption{\added{$K_S-$WISE~4 color versus $J-K_S$ color for the comparison and reddened stars in our sample. The horizontal dotted line at $K_S-\text{WISE4} = 1$ divides between sources with and without strong stellar winds. The red circles represent reddened stars that were not used to measure extinction curves (see Fig.~\ref{fig:bad_stars}), because they have winds. The purple square corresponds to \object{HD014250}, which was not used because it has other peculiarities in its spectrum (as explained in Section \ref{sec:measure}). The green diamonds represent the reddened stars that were used to measure extinction curves. They do not show signatures of stellar winds. The green diamond above the line corresponds to \object{HD029647}, which is not windy (as confirmed by \cite{2021ApJ...916...33G}), but its WISE photometry possibly suffers from extended source and/or scattered moonlight contamination.} \label{fig:wind}}
\end{figure}

The normalized extinction curves are shown in Fig.~\ref{fig:ext_curves}\added{, and are electronically \dataset[available]{\doi{10.5281/zenodo.5802469}} \citep{decleir_marjorie_2022_5802469}}. We were able to measure extinction curves for 15 reddened stars of our sample (with spectra shown in Fig.~\ref{fig:red_stars}). The other 10 reddened stars were not suitable to measure extinction curves, due to peculiarities in their spectrum (plotted in Fig.~\ref{fig:bad_stars}). Four of these stars (\object{HD014422}, \object{HD034921}, \object{HD206773} and \object{HD052721}) show strong emission lines in their spectrum (as also indicated in their spectral type with the letter ``e"). \object{HD166734} and \object{HD294264} have a strongly rising spectrum towards longer wavelengths. Emission lines and/or very steeply rising spectra are most likely signatures of a stellar wind or circumstellar disk. \added{\object{HD034921}, \object{HD206773} and \object{HD166734} have also been classified as stars with clear wind signatures by \cite{2021ApJ...916...33G}. As an additional check, we used a similar approach as in Fig.~4 of \cite{2021ApJ...916...33G} to separate stars with wind signatures, based on their IR colors. Fig.~\ref{fig:wind} shows $K_S-$WISE~4 versus $J-K_S$ for the reddened and comparison stars in our sample that have WISE photometry \citep{2010AJ....140.1868W}, obtained from the \dataset[IRSA AllWISE Source Catalog]{\doi{10.26131/IRSA1}} \citep{irsawise}. The $J-K_S$ color is a measure of the reddening and high $K_S-$WISE~4 values ($\gtrsim1$) indicate stellar wind signatures.} The 4 remaining stars (\object{HD037020}, \object{HD037023}, \object{HD037022} and \object{HD014250}) have either an ``n" (indicating nebulous absorption) or a ``p" (indicating another unspecified peculiarity) in their spectral type. All these peculiarities are unique to the specific star and complicate the extinction measurement using the pair method, because a comparison star with the same characteristics (e.g. same wind emission) would be needed. We thus excluded these sightlines from all further analyses. \added{We indicated with an * in Table~\ref{tab:sample} which stars we used to measure extinction curves. These are plotted as green diamonds in Fig.~\ref{fig:wind}, and do not show signatures of stellar winds.}

\added{The median SNRs of the normalized extinction curves are in the range 2.7--27 below 2.5\,\micron, 0.4--5.8 between 2.8 and 4\,\micron, and 0.3--1.5 beyond 4.5\,\micron, at a resolution of about 2000.} We would like to note here that some of the 2MASS photometric data points in Fig.~\ref{fig:ext_curves} seem to deviate slightly from the spectral extinction curves. This is likely due to saturation issues in the J-band for some of the (bright) stars in our sample, resulting in a lower quality (as flagged in the 2MASS catalog) and hence in larger error bars on the magnitudes, and consequently on the extinctions in those bands. Given that all three bands (J, H and K\textsubscript{S}) were used to scale the (SXD) spectrum (as discussed in Section~\ref{sec:spectra}), a small offset in one of the bands could slightly shift the spectrum up or down, causing the other bands to look slightly deviant from the spectrum as well. However, because we are using all three bands for the scaling, an issue with one band does not significantly impact the extinction measurements and fitting results. Furthermore, all JHK\textsubscript{S} photometric data points (for all stars) are within 3$\sigma$ from the spectrum. Finally, when available, we tested with other JHK photometry, but this did not significantly affect the final fitting results. Hence, we decided to stick with the 2MASS photometry for all sightlines to be as uniform as possible.

\section{Fitting NIR extinction curves}
\label{sec:fitting}

\subsection{Continuum extinction}
\label{sec:continuum}
In previous works, the continuum NIR extinction was usually described with a power law \citep[see e.g.,][]{1989ESASP.290...93D, 1989ApJ...345..245C, 1990ApJ...357..113M}, as it provides a convenient analytical representation of the extinction curve. We followed their example and fitted the continuum extinction between 0.8 and 5.5\,\micron\ with a power law. From Eq.~\ref{eq:reddening}, the measured differential extinction can be written as:
\begin{equation}
\label{eq:elv}
E(\lambda - V) = A(V) \left[ k(\lambda) - 1 \right]
\end{equation}
where
\begin{equation}
\label{eq:powerlaw}
k(\lambda) = \frac{A(\lambda)}{A(V)}  =  S\,\lambda^{-\alpha}
\end{equation}
where $S$ is the amplitude (by definition equal to the normalized extinction at 1\,\micron) and $\alpha$ is the index of the power law. At infinite wavelengths, Eq.~\ref{eq:elv} reduces to $E(\infty-V) = -A(V)$. In other words, fitting the observed extinction curve with the function in Eqs.~\ref{eq:elv} and \ref{eq:powerlaw}, gives us a direct measurement of $A(V)$. We used a combination of the \texttt{PowerLaw1D} Astropy \citep{2013A&A...558A..33A} model (for Eq.~\ref{eq:powerlaw}), and the \texttt{AxAvToExv} conversion function from the \texttt{dust\_extinction} python package \citep{dustextinction} to implement the conversion in Eq.~\ref{eq:elv}. 

The fitting was done in two steps. First, the Levenberg–Marquardt algorithm was used to obtain preliminary fit results for the three parameters $S$, $\alpha$ and $A(V)$, using the \texttt{LevMarLSQFitter} from Astropy. These results were then used as initial guesses for the Markov chain Monte Carlo (MCMC) fitting based on the \texttt{Emcee} python tool \citep{2013PASP..125..306F}. We used 6 walkers, each with 9000 steps after a burn in of 1000 steps to sample the parameter space. We experimented with more walkers, more steps and larger burn fractions, but the results were the same. \added{The inverse of the squared uncertainties on the extinction were used as weights in the fitting, so that wavelength regions with larger uncertainties contribute less to the fit. The fitted curves are shown as red dashed lines in Fig.~\ref{fig:ext_curves}.}

The uncertainties on the fitted parameters given by the MCMC fitting only include random noise, computed as the difference between the 84\textsuperscript{th} and 50\textsuperscript{th} percentile (upper uncertainty), or between the 50\textsuperscript{th} and 16\textsuperscript{th} percentile (lower uncertainty) of the posterior distribution function. However, there is also a systematic uncertainty related to the fact that the extinction curves are measured relative to the V-band extinction. To assess the effect of this uncertainty, we first combined the uncertainties on the V-band magnitude of the reddened and the comparison star by adding them quadratically and taking the square root. Subsequently, we ran the MCMC fitting three times: using the measured extinction, using the extinction subtracted by the combined V-band uncertainty, and using the extinction summed with the combined V-band uncertainty. For all parameters, we then computed the systematic uncertainty as half of the difference between the maximum and minimum of the three fitted values. Finally, we added this systematic uncertainty in quadrature to the random uncertainty\added{, and took the square root to obtain the total uncertainty on each fitted value}.

The code that was used to perform the fitting, compute the uncertainties, and analyze and plot the results, is available as part of the \texttt{spex\_nir\_extinction} package on GitHub \citep{marjorie_decleir_2022_6330037}.

\vfill

\subsection{Extinction features}
\label{sec:features}
Strong extinction features can influence the shape of the fitted power law, because the procedure described above will try to fit the features and the continuum together with one power law. Therefore, for sightlines with strong features, it is important to explicitly consider these features in the fitting by, e.g., adding a Drude profile to the power law, and fit the data with the combined function, as such constraining the free parameters of the power law and the Drude function simultaneously. 

\begin{figure*}
\gridline{\leftfig{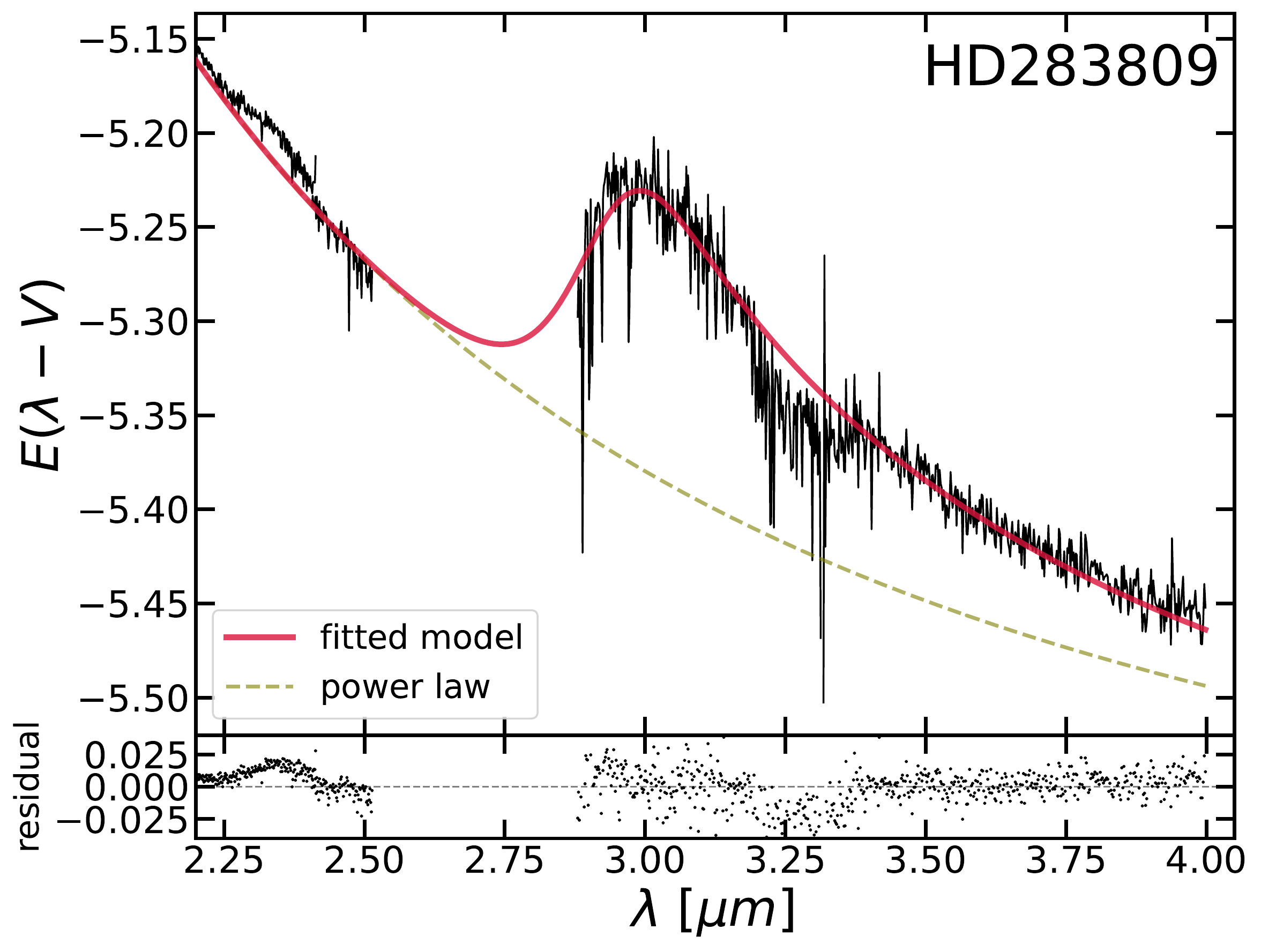}{0.49\textwidth}{}
          \rightfig{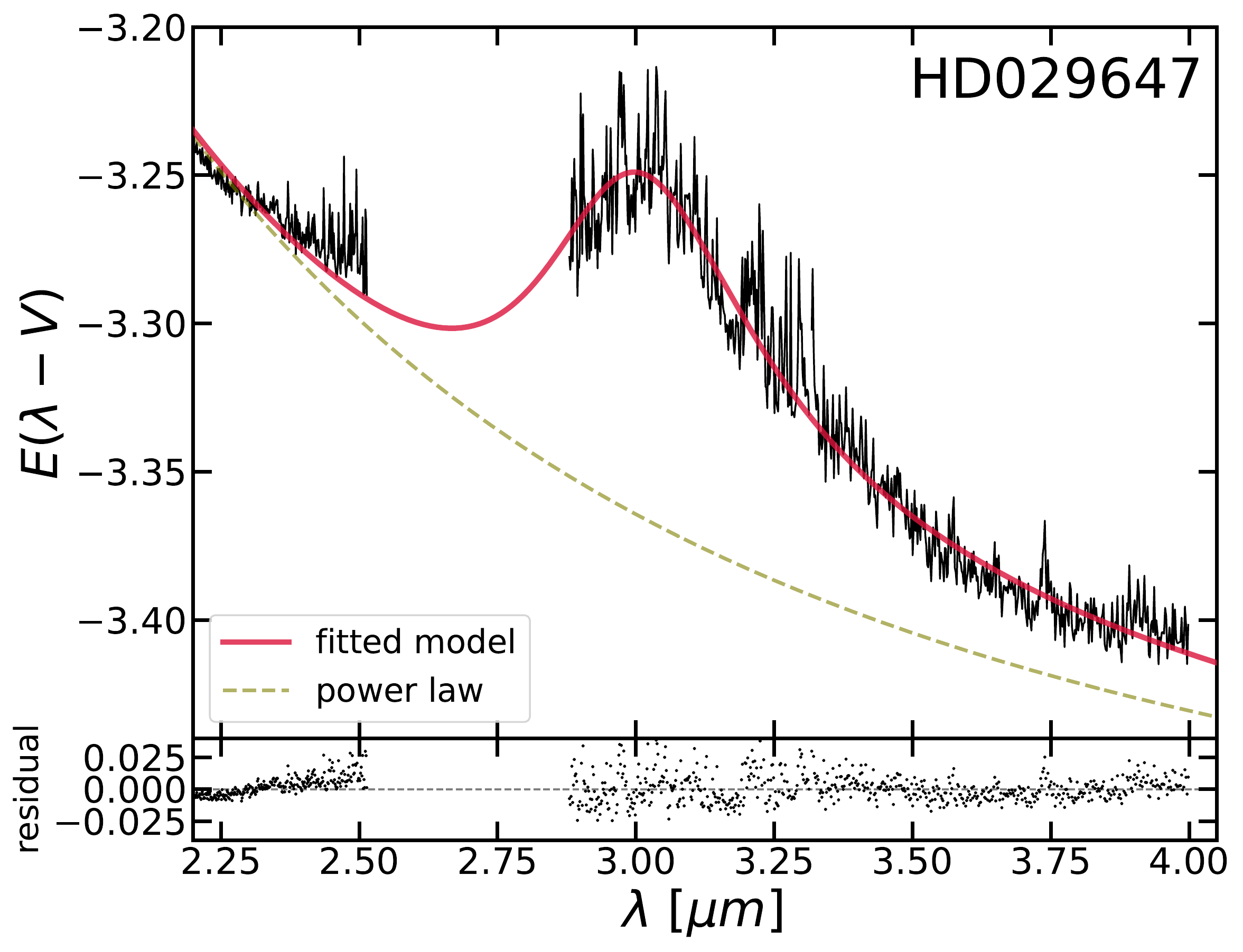}{0.49\textwidth}{}
          }
\caption{The extinction curves of the sightlines towards \object{HD283809} (left) and \object{HD029647} (right) show a strong ice feature around 3\,\micron. The continuum extinction and the ice feature were fitted simultaneously with a linear combination of a power law and a modified Drude profile. The fit parameters are given in Table~\ref{tab:fit_results_dense}. \label{fig:features}}
\end{figure*}

From a visual inspection of the extinction curves, only two sightlines in our sample seem to have an obvious extinction feature around 3\,\micron: \object{HD283809} and \object{HD029647}, as can be seen in Fig.~\ref{fig:features}. This feature is caused by water ice along the line-of-sight, as will be discussed in more detail in Section \ref{sec:features_lab}. Since it is generally accepted that the detection of ice indicates the presence of a dense molecular component along the line-of-sight \citep[e.g.,][]{1997ApJ...490..729W,2015ARA&A..53..541B}, we consider these two sightlines as ``dense''. Note that we did not use the total V-band extinction $A(V)$ to divide our sample into diffuse and dense sightlines, as it is known that a high $A(V)$ can also be measured in sightlines without any dense material (e.g., in the diffuse sightline towards Cyg~OB2~no.\,12 which has $A(V)\sim10$, \cite{2015ApJ...811..110W}). The extinction curve of \object{HD283809} also shows a weaker bump around 3.4\,\micron, which is likely caused by hydrocarbons, as will be discussed in Section \ref{sec:features_lab}. This feature is, however, not obvious in the extinction curve towards \object{HD029647}. There also seems to be a small systematic residual around 2.3\,\micron\ for \object{HD283809}, which is not visible for \object{HD029647}. However, it is possible that this residual is due to the\deleted{ less smooth} transition between the SXD and LXD spectrum around 2.4\,\micron.

One approach to study these extinction features would be to compare them with laboratory measurements, as was e.g. done by \cite{2006A&A...449..251T} and \cite{2011ApJ...729...92B}. However, since \replaced{the main goal}{one of the goals} of this work is to obtain a functional form for the NIR extinction curve, we opted to directly fit functional profiles to our data, without making any assumptions on the material causing the features. \deleted{A brief discussion on the identification of these features is given in Section \ref{sec:features_lab}.} 

It has to be noted here that it is also possible to constrain an extinction feature directly from the spectrum, as has been done in the past \citep[e.g.,][]{2002ApJS..138...75P,2006A&A...449..251T}. However, this requires knowledge of the underlying continuum flux, which is often approximated by a straight line between two points left and right from the feature. Similarly, fitting extinction features from a continuum-subtracted extinction curve requires assumptions on the underlying continuum extinction and the extent of the features, which introduces uncertainties on the measurement of the extinction features \citep[e.g.,][]{2011ApJ...729...92B}. Therefore, we have opted to fit the extinction continuum and features simultaneously \added{in the extinction curves}.

The strong telluric absorption (see Fig.~\ref{fig:atmos}) significantly increases the noise between 2.5 and 3.5\,\micron, which makes it hard to obtain any meaningful constraints on the weaker feature around 3.4\,\micron. Furthermore, we found that the impact of this weak bump on the continuum fit is negligible. We thus only considered the strong ice feature in the fitting. We fitted the extinction curve \added{of both dense sightlines} with a combination of a power law and a modified Drude profile $D(\lambda)$:
\begin{equation}
\label{eq:full}
k(\lambda) = \frac{A(\lambda)}{A(V)}  =  S\,\lambda^{-\alpha} + B\, D(\lambda)
\end{equation}

\noindent with
\begin{equation}
D(\lambda)= \frac{(\gamma/\lambda_0)^2}{(\lambda/\lambda_0-\lambda_0/\lambda)^2+(\gamma/\lambda_0)^2}
\label{eq:drude}
\end{equation}

\noindent with $B$ the strength, $\lambda_0$ the central wavelength, and $\gamma$ the width of the profile. Because of the asymmetric appearance of the feature, we modified the standard Drude function to allow for extra asymmetry.\footnote{Note that a Drude profile is intrinsically already asymmetric, but our modified version allows for extra asymmetry by implementing an \textit{asymmetry} parameter.} The modified profile is obtained by replacing the standard width in Eq.~\ref{eq:drude} by a width $\gamma(\lambda)$ that depends on the wavelength:
\begin{equation}
\gamma(\lambda) = \frac{2 \gamma_0}{1 + \text{exp}[a(\lambda - \lambda_0)]}
\end{equation}
with $\gamma_0$ the standard width, and $a$ the asymmetry parameter.
This modification is based on the work of \citet{STANCIK200866}, who introduced asymmetric Gaussian and Lorentzian profiles with widths varying across the profiles to model infrared absorption profiles. It was also successfully used by \cite{2021ApJ...916...33G} to fit the 10 and 20\,\micron\ silicate features in MIR extinction curves.

The fitting results \added{(i.e. the 50\textsuperscript{th} percentiles of the posterior distribution functions) and their uncertainties (including both random and systematic uncertainties, as explained in Section~\ref{sec:continuum})} for the two dense sightlines are given in Table~\ref{tab:fit_results_dense}. To verify whether an ice feature is present in the extinction curves of the other sightlines, we also fitted them with a combination of a power law and a modified Drude profile as described above, but with the central wavelength, width and asymmetry fixed to the average values of the two dense sightlines. For only 4 sightlines, the fitted strength of the Drude profile is larger than 3 times its uncertainty (i.e. a $>3\sigma$ detection): \object{HD038087}, \object{HD156247}, \object{HD183143} and \object{HD229238}. However, an investigation of the extinction curves and residuals of \object{HD038087} and \object{HD156247}, shows that these sightlines significantly suffer from telluric absorption around 3\,\micron. In sightlines \object{HD183143} and \object{HD229238}, there is a very tentative detection of a weak ice feature with strengths $B=0.0026\pm0.0008$ ($3.3\sigma$) and $B=0.0042\pm0.0012$ ($3.5\sigma$) respectively (as compared to a $>50\sigma$ detection for the two dense sightlines). These sightlines have the highest $A(V)$ values of the diffuse sample ($A(V)=3.86$ and $A(V)=2.99$, respectively). Given the non-detection (or very tentative detection) of ice in all sightlines \added{other than \object{HD283809} and \object{HD029647}}, \replaced{they}{these} can be considered ``diffuse" for the purpose of this paper. The fitting results \added{and their uncertainties} for the 13 diffuse sightlines (without fitting any features) are given in Table~\ref{tab:fit_results_diff}.

\begin{deluxetable*}{ll|CCCCC}
\tablecaption{MCMC fitting results for the 13 diffuse extinction curves and the average diffuse extinction curve\deleted{: the amplitude $S$ and index $\alpha$ of the power law, and $A(V)$ are directly obtained from the fitting (50\textsuperscript{th} percentile of the posterior distribution function), while $E(B-V)$ is obtained from the observations, and $R(V)$ is calculated as $R(V)=A(V)/E(B-V)$. The reported uncertainties include both random and systematic uncertainties, as explained in Section \ref{sec:continuum}}. \label{tab:fit_results_diff}}
\tablehead{\colhead{reddened} & \colhead{comparison} & \colhead{$S$} & \colhead{$\alpha$} & \colhead{$A(V)$} & \colhead{$E(B-V)$} & \colhead{$R(V)$}}
\startdata
\object[BD +56 524]{BD+56d524} & \object{HD034816} & 0.402_{-0.005}^{+0.005} & 1.58_{-0.02}^{+0.02} & 1.78_{-0.02}^{+0.02} & 0.59_{-0.04}^{+0.04} & 3.01_{-0.19}^{+0.20} \\
\object{HD013338} & \object{HD031726} & 0.435_{-0.004}^{+0.004} & 1.69_{-0.02}^{+0.02} & 1.38_{-0.01}^{+0.01} & 0.45_{-0.01}^{+0.01} & 3.10_{-0.10}^{+0.11} \\
\object{HD014956} & \object{HD214680} & 0.404_{-0.004}^{+0.004} & 1.44_{-0.01}^{+0.01} & 2.82_{-0.03}^{+0.03} & 0.92_{-0.05}^{+0.05} & 3.07_{-0.16}^{+0.17} \\
\object{HD017505} & \object{HD214680} & 0.210_{-0.003}^{+0.003} & 2.20_{-0.03}^{+0.03} & 1.51_{-0.02}^{+0.02} & 0.60_{-0.04}^{+0.04} & 2.51_{-0.15}^{+0.17} \\
\object{HD029309} & \object{HD042560} & 0.484_{-0.008}^{+0.008} & 1.42_{-0.01}^{+0.01} & 2.02_{-0.03}^{+0.03} & 0.49_{-0.05}^{+0.05} & 4.12_{-0.36}^{+0.43} \\
\object{HD037061} & \object{HD034816} & 0.458_{-0.004}^{+0.004} & 1.36_{-0.01}^{+0.01} & 2.52_{-0.02}^{+0.02} & 0.51_{-0.04}^{+0.04} & 4.94_{-0.34}^{+0.40} \\
\object{HD038087} & \object{HD051283} & 0.445_{-0.007}^{+0.007} & 1.66_{-0.01}^{+0.01} & 1.55_{-0.03}^{+0.03} & 0.29_{-0.04}^{+0.04} & 5.33_{-0.64}^{+0.85} \\
\object{HD156247} & \object{HD042560} & 0.403_{-0.026}^{+0.026} & 1.93_{-0.06}^{+0.06} & 0.78_{-0.05}^{+0.05} & 0.22_{-0.07}^{+0.07} & 3.52_{-0.88}^{+1.67} \\
\object{HD183143} & \object{HD188209} & 0.367_{-0.002}^{+0.002} & 1.90_{-0.01}^{+0.01} & 3.86_{-0.02}^{+0.02} & 1.29_{-0.04}^{+0.04} & 2.99_{-0.09}^{+0.09} \\
\object{HD185418} & \object{HD034816} & 0.325_{-0.006}^{+0.006} & 2.02_{-0.03}^{+0.03} & 1.23_{-0.02}^{+0.02} & 0.47_{-0.04}^{+0.04} & 2.61_{-0.20}^{+0.23} \\
\object{HD192660} & \object{HD214680} & 0.391_{-0.006}^{+0.006} & 1.71_{-0.01}^{+0.01} & 2.31_{-0.04}^{+0.04} & 0.87_{-0.06}^{+0.06} & 2.65_{-0.18}^{+0.21} \\
\object{HD204827} & \object{HD003360} & 0.340_{-0.005}^{+0.005} & 2.08_{-0.01}^{+0.01} & 2.43_{-0.03}^{+0.03} & 1.00_{-0.05}^{+0.05} & 2.43_{-0.11}^{+0.12} \\
\object{HD229238} & \object{HD214680} & 0.340_{-0.003}^{+0.003} & 1.99_{-0.01}^{+0.01} & 2.99_{-0.02}^{+0.02} & 1.10_{-0.04}^{+0.04} & 2.71_{-0.09}^{+0.10} \\
\hline
\multicolumn{2}{c|}{average diffuse} & 0.386_{-0.001}^{+0.001}\tablenotemark{a} & 1.71_{-0.01}^{+0.01}\tablenotemark{a} &  &  & 3.12$\pm$0.05\tablenotemark{b}\\
\enddata
\tablenotetext{a}{Obtained by fitting the average diffuse extinction curve (see Section \ref{sec:ave_fit}).}
\tablenotetext{b}{Calculated as $R(V)=\frac{1}{\left(\frac{A(B)}{A(V)}-1\right)}$.}
\tablecomments{\added{The measured extinction values in this table are relative to the comparison star. The reported uncertainties include both random and systematic uncertainties, as explained in Section \ref{sec:continuum}.}}
\end{deluxetable*}

\begin{deluxetable*}{ll|CCCCCCCCC}
\tabletypesize{\footnotesize}
\tablecaption{MCMC fitting results for the 2 dense extinction curves\deleted{: the amplitude $S$ and index $\alpha$ of the power law, the strength $B$, central wavelength $\lambda_0$, width $\gamma_0$ and asymmetry $a$ of the Drude profile, and $A(V)$ are directly obtained from the fitting (50\textsuperscript{th} percentile of the posterior distribution function), while $E(B-V)$ is obtained from the observations, and $R(V)$ is calculated as $R(V)=A(V)/E(B-V)$. The reported uncertainties include both random and systematic uncertainties, as explained in Section \ref{sec:continuum}}. \label{tab:fit_results_dense}}
\tablehead{\colhead{reddened} & \colhead{comparison} & \colhead{$S$} & \colhead{$\alpha$} & \colhead{$B$} & \colhead{$\lambda_0$} & \colhead{$\gamma_0$} & \colhead{$a$} & \colhead{$A(V)$} & \colhead{$E(B-V)$} & \colhead{$R(V)$}}
\startdata
\object{HD029647} & \object{HD034759} & 0.369_{-0.003}^{+0.003} & 1.93_{-0.01}^{+0.01} & 0.0331_{-0.0006}^{+0.0007} & 3.022_{-0.005}^{+0.005} & 0.49_{-0.03}^{+0.03} & -1.32_{-0.22}^{+0.19} & 3.52_{-0.03}^{+0.03} & 1.05_{-0.04}^{+0.04} & 3.35_{-0.13}^{+0.14} \\
\object{HD283809} & \object{HD003360} & 0.389_{-0.002}^{+0.002} & 1.91_{-0.01}^{+0.01} & 0.0264_{-0.0005}^{+0.0005} & 3.014_{-0.006}^{+0.005} & 0.44_{-0.02}^{+0.02} & -4.38_{-0.83}^{+0.64} & 5.65_{-0.03}^{+0.03} & 1.61_{-0.06}^{+0.06} & 3.51_{-0.12}^{+0.13}
\enddata
\tablecomments{\added{The measured extinction values in this table are relative to the comparison star. The reported uncertainties include both random and systematic uncertainties, as explained in Section \ref{sec:continuum}.}}
\end{deluxetable*}

\section{Results and discussion} \label{sec:results}

\subsection{Extinction curve parameter trends}
\label{sec:trends}
\begin{figure}
\epsscale{1.1}
\plotone{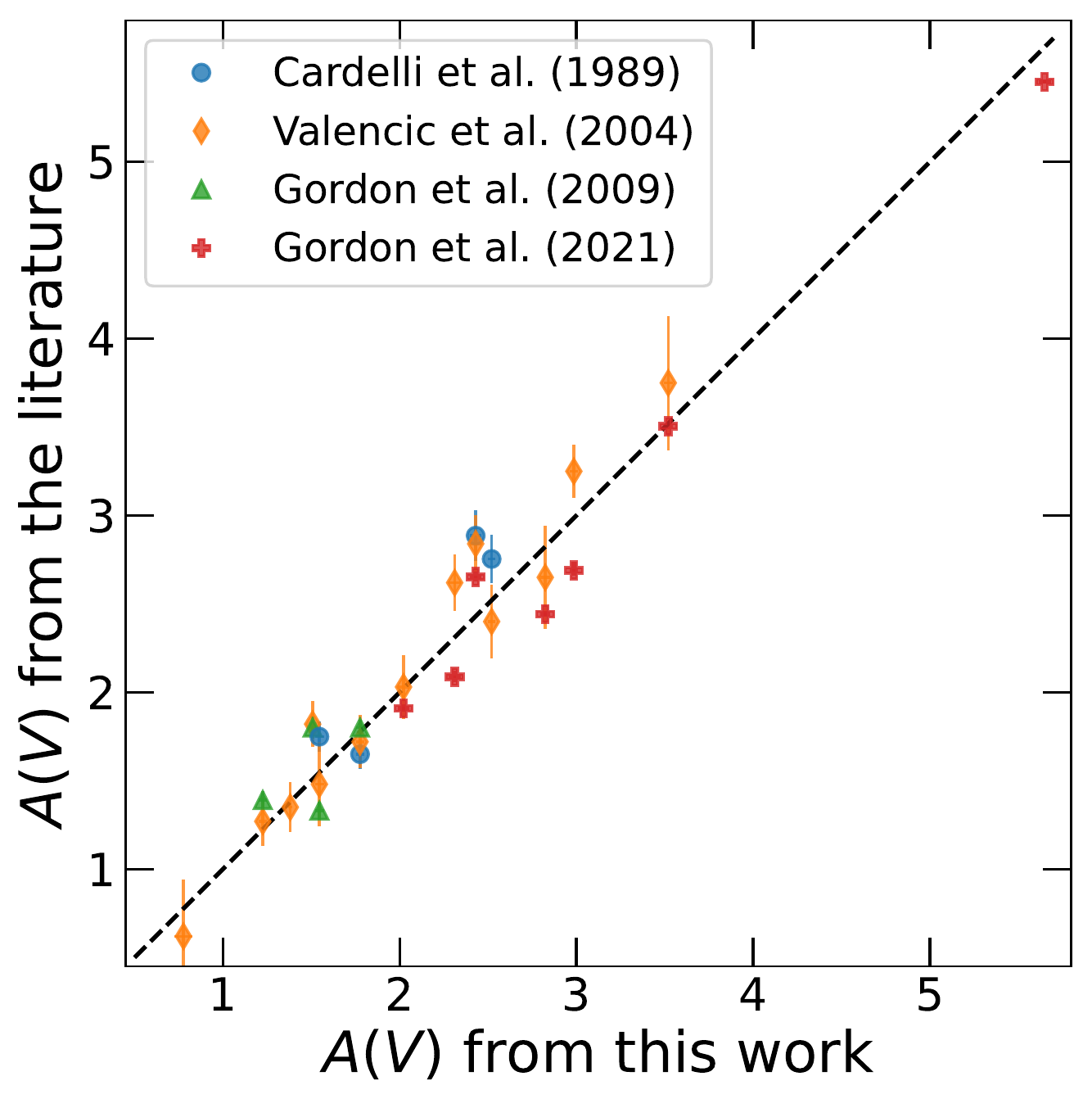}
\caption{A comparison between our obtained $A(V)$ values from the fitting (see Tables~\ref{tab:fit_results_diff} and \ref{tab:fit_results_dense}) to values reported in the literature for the same sightlines (taken from \citealt{1989ApJ...345..245C}, \citealt{2004ApJ...616..912V}, \citealt{2009ApJ...705.1320G} and \citealt{2021ApJ...916...33G}.). There is a good agreement for all sightlines. \label{fig:AV_comp}}
\end{figure}

The fitting results in Tables~\ref{tab:fit_results_diff} and~\ref{tab:fit_results_dense} show a wide range in $A(V)$ (0.78--5.65) and $R(V)$ values (\replaced{2.33}{2.43}--5.33).
As a cross-check, in Fig.~\ref{fig:AV_comp} we compared our obtained $A(V)$ values with values reported in the literature for the same sightlines. $A(V)$ values were collected from \cite{1989ApJ...345..245C}, \cite{2004ApJ...616..912V}, \cite{2009ApJ...705.1320G} and \cite{2021ApJ...916...33G}\added{, which all used the pair method to measure extinction curves. 
\cite{1989ApJ...345..245C} fitted the \cite{1985ApJ...288..618R} extinction curve to the R, I, J, H, K, and L color excesses $E(\lambda-V)$ with $A(V)$ as the only free parameter. \cite{2004ApJ...616..912V} estimated $R(V)$ from the J, H, and K color excesses as described in \cite{1999PASP..111...63F}, and calculated $A(V)$ using $R(V)$ and $E(B-V)$. \cite{2009ApJ...705.1320G} extrapolated the J, H, and K $E(\lambda-V)$ curves to infinite wavelength using the \cite{1989ApJ...336..752R} IR extinction curve to derive $A(V)$ values. Finally, \cite{2021ApJ...916...33G} derived $A(V)$ by fitting a combination of a power law and two modified Drude functions (for the silicate features) to their measured NIR/MIR $E(\lambda-V)$ extinction curves.} The agreement \added{between our $A(V)$ values and the literature values} is very good for all sightlines\added{, even though they were derived using different methods and different data}.

\begin{figure*}[ht]
\centering
\includegraphics[width=0.75\textwidth]{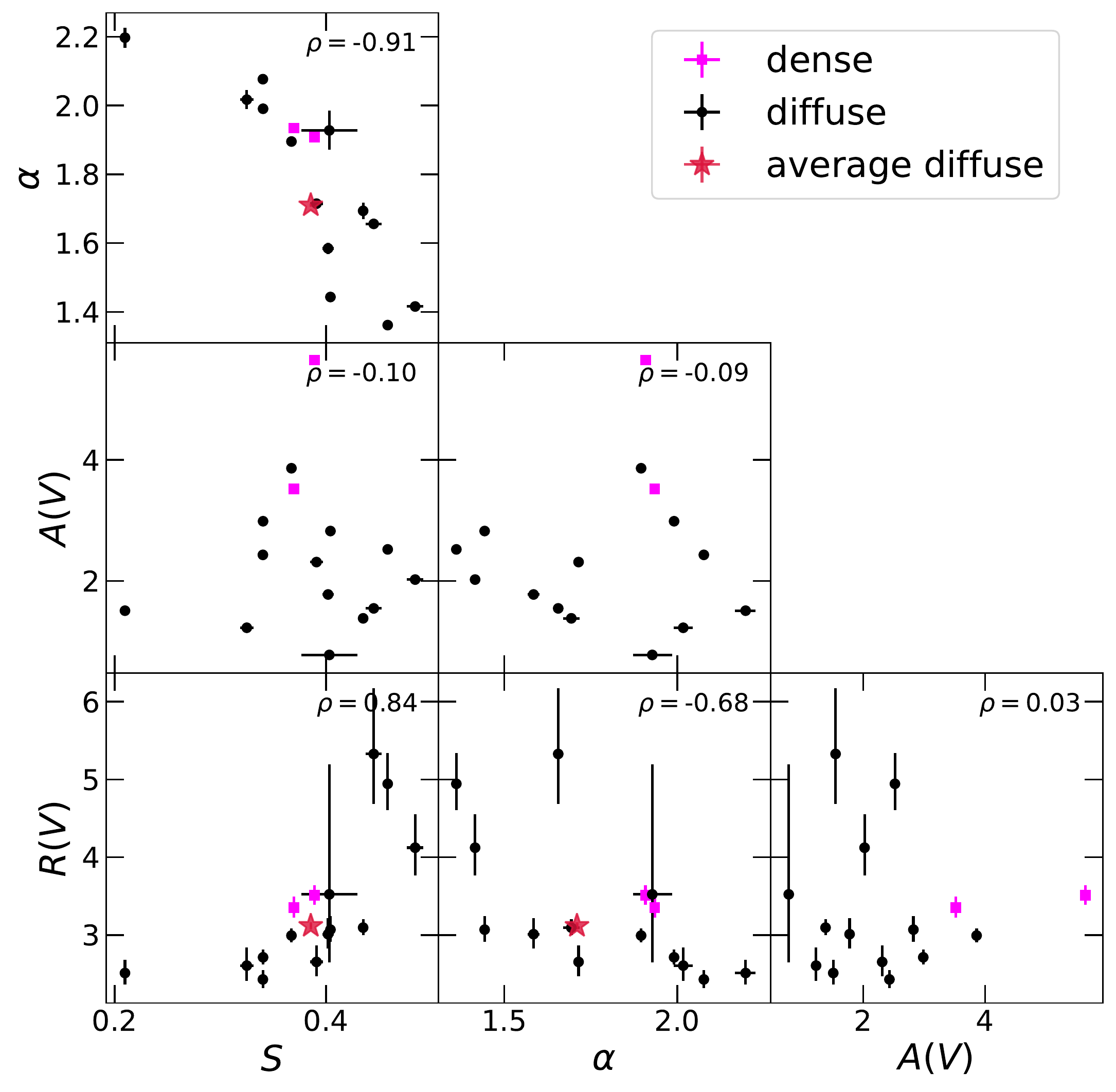}
\caption{Scatter plots of the different parameters describing the extinction curve: the amplitude $S$ and the index $\alpha$ of the  power law, the V-band extinction $A(V)$ and the total-to-selective extinction $R(V)$ (see Tables~\ref{tab:fit_results_diff} and \ref{tab:fit_results_dense}). The magenta squares indicate the dense sightlines, the black circles indicate the diffuse sightlines, and the red star shows the average diffuse extinction curve (see Section \ref{sec:ave_fit}). The Spearman's rank correlation coefficient is shown in every plot. \label{fig:params}}
\end{figure*}

It is interesting to see if there are any correlations between the different parameters describing the extinction curve: the amplitude $S$ and the index $\alpha$ of the power law, the V-band extinction $A(V)$ and the total-to-selective extinction $R(V)$. Fig.~\ref{fig:params} shows scatter plots of these parameters. We found a clear anti-correlation (with Spearman's rank correlation coefficient $\rho=-0.91$) between the amplitude and the index of the power law. This is straightforward to explain considering the fact that the amplitude is equal to the normalized extinction at 1\,\micron, and that the extinction curves are normalized to $A(V)$. A steeper extinction curve (i.e. larger $\alpha$) implies a faster decrease in extinction at wavelengths beyond the V-band compared to a flatter extinction curve, resulting in a lower normalized extinction at 1\,\micron\ (i.e. smaller $S$). $R(V)$ correlates with the amplitude of the power law (\replaced{$\rho=0.77$}{$\rho=0.84$}), and anti-correlates with the power law index (\replaced{$\rho=-0.61$}{$\rho=-0.68$}). This shows that $R(V)$ is linked to the slope of the NIR extinction curve, with higher $R(V)$ values corresponding to flatter curves (i.e. smaller $\alpha$ and larger $S$), as is also the case in the UV and optical \citep[e.g.,][]{1989ApJ...345..245C,1999PASP..111...63F}. The $R(V)$-dependence of the shape of the extinction curve is investigated and discussed in more detail in Section \ref{sec:RV_dep}. No correlation was found between $A(V)$ and any of the other parameters ($\rho=-0.10$, $\rho=-0.09$ and \replaced{$\rho=0.00$}{$\rho=0.03$}). 

\subsection{Average diffuse NIR Milky Way extinction curve}
\label{sec:average}

\added{\subsubsection{Measurement and fitting}\label{sec:ave_fit}}

\begin{figure*}
\gridline{\leftfig{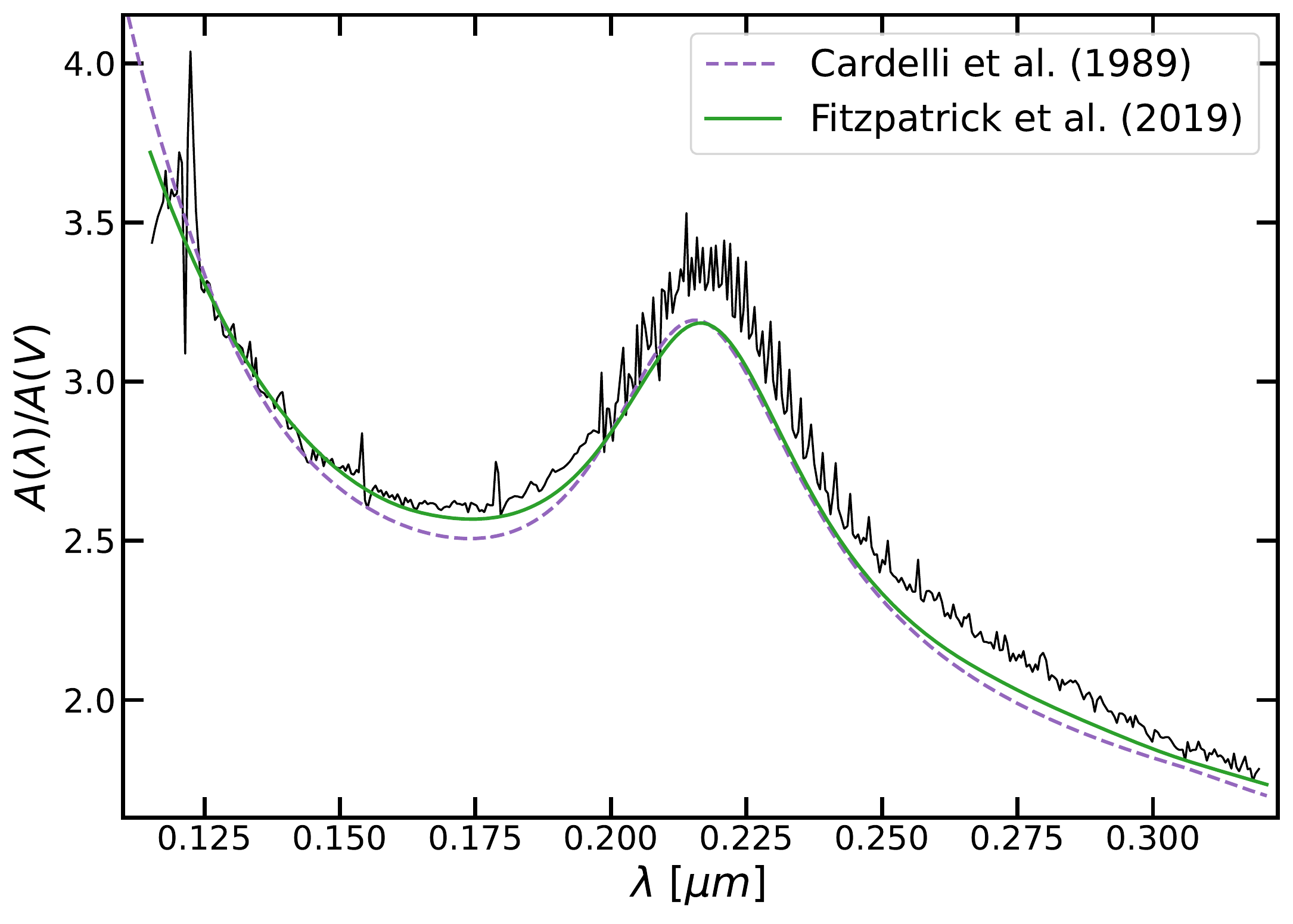}{0.496\textwidth}{}
          \rightfig{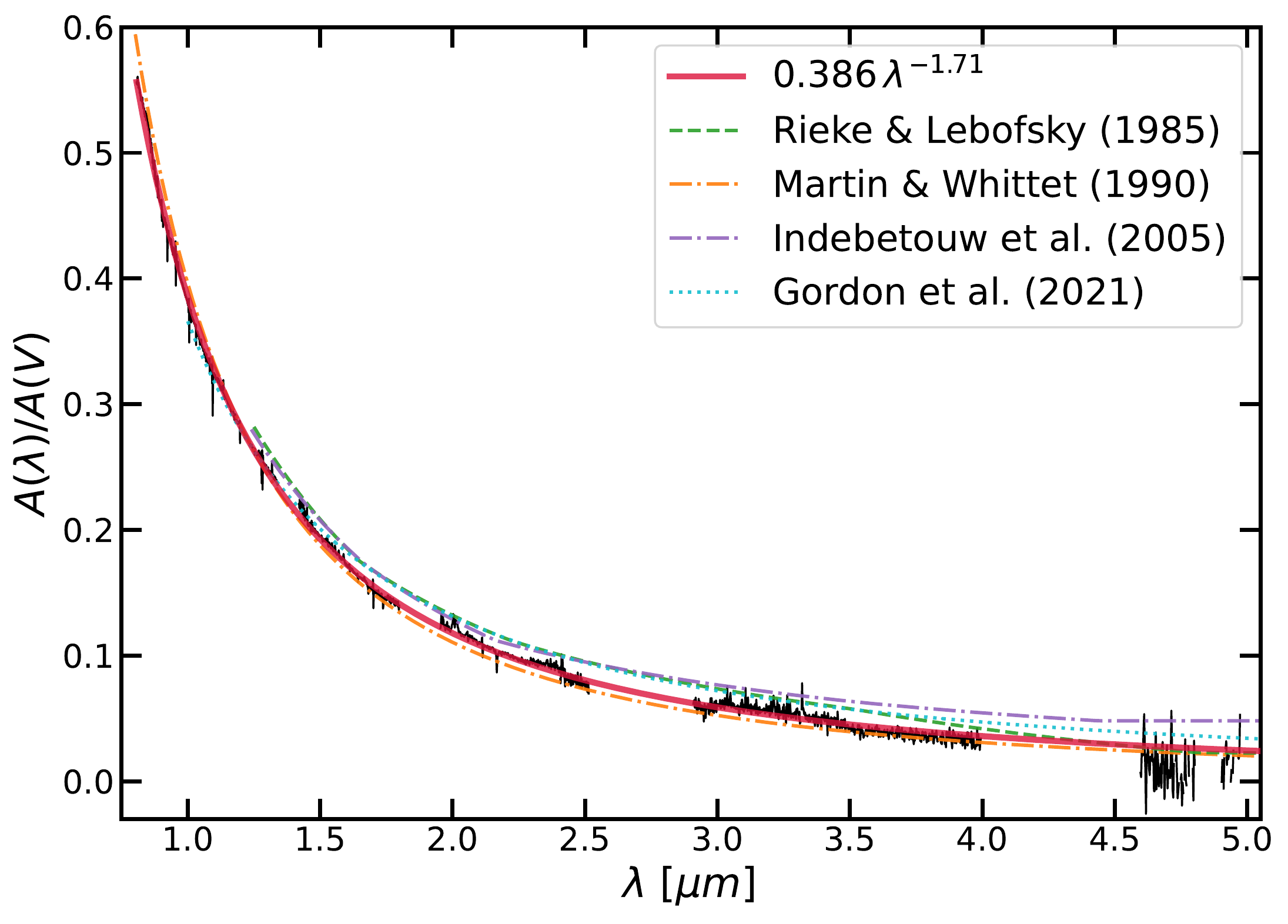}{0.496\textwidth}{}
          }
\caption{Left: Average diffuse UV extinction curve \added{(measured from IUE spectra \citep{2004ApJ...616..912V} and} rebinned by a factor of 3 to reduce the noise), together with average \added{diffuse} curves from the literature: \cite{1989ApJ...345..245C} and \cite{2019ApJ...886..108F}. Right: Average diffuse NIR extinction curve, together with average curves from the literature: \cite{1985ApJ...288..618R}, \cite{1990ApJ...357..113M}, \cite{2005ApJ...619..931I}, and \cite{2021ApJ...916...33G}. The red line shows the power law fit of the average curve.
\label{fig:ave}}
\end{figure*}

We averaged the extinction curves of the 13 diffuse sightlines in our sample. The average was only computed in those wavelength regions where at least 5 curves have data, as such excluding the wavelengths with limited and noisy extinction data (e.g. beyond 5\,\micron). The average NIR curve is plotted in the right panel of Fig.~\ref{fig:ave}, together with \replaced{several}{some} average Milky Way extinction curves from the literature \citep{1985ApJ...288..618R, 1990ApJ...357..113M,2005ApJ...619..931I,2021ApJ...916...33G}. Our average curve is very close to that from \cite{1990ApJ...357..113M}, while the other literature curves are slightly flatter. Note that all these average curves are based on a different sample of sightlines and some differences are to be expected. We also compared with average curves from \cite{1989ApJ...336..752R} and \cite{2006ApJ...637..774C}, which are not shown in the plot because they are very similar to the other literature curves. \added{A more detailed comparison to previous studies is given in Section~\ref{sec:comparison}.} Given that our average curve is close to most \added{diffuse} literature curves, we can conclude that our sample of 13 diffuse sightlines provides a reasonable representation of the average diffuse Milky Way extinction. As an additional check, we also measured the average UV extinction curve for our sample of sightlines using International Ultraviolet Explorer (IUE) \dataset[spectra]{10.17909/165k-w223} \citep[taken from][]{2004ApJ...616..912V}. The average UV curve for our sample is plotted in the left panel of Fig.~\ref{fig:ave}, and is very similar in shape to the \cite{1989ApJ...345..245C} and \cite{2019ApJ...886..108F} average \added{$R(V) = 3.1$ diffuse} UV curves, hence confirming that our sample of 13 diffuse sightlines is representative of the average diffuse Milky Way extinction. \added{Furthermore, we find $R(V)=3.12\pm0.05$ for our average curve, which is consistent with the average diffuse Milky Way value \citep{1989ApJ...345..245C}.} \deleted{The peaks in the UV extinction curve show that there are small spectral mismatches between some of the reddened stars and their comparison star, which are less obvious in the NIR.} \replaced{A binned version of the measured average NIR extinction curve}{The measured average NIR extinction curve is electronically \dataset[available]{\doi{10.5281/zenodo.5802469}} \citep{decleir_marjorie_2022_5802469}, and a binned version of the curve} and its uncertainty is given in Table~\ref{tab:RV_dep} (columns 2 and 3) for wavelengths between 0.8 and 4\,\micron, and in common IR photometric bands (2MASS J, H, K\textsubscript{S}, WISE~1, L, and IRAC~1), obtained by integrating the average extinction curve over the response curves of the bands (taken from the SVO Filter Profile Service -- \cite{2012ivoa.rept.1015R, 2020sea..confE.182R}).

\startlongtable
\begin{deluxetable}{c|ccc|ccc}
\tablecaption{Average diffuse Milky Way extinction curve and parameters of the linear relationship between extinction $A(\lambda)/A(V)$ and $1/R(V)$. This table is also available as part of the D22\_MWAvg model in the \texttt{dust\_extinction} python package \citep{dustextinction}. \label{tab:RV_dep}}
\tablehead{\colhead{band} & \multicolumn3c{average extinction} & \multicolumn3c{$R(V)$-dependent ext.}\\
\colhead{or} & \multicolumn{2}{c}{measured} & \colhead{fit} & \nocolhead{}\\
\colhead{$\lambda\ [\micron]$} & \colhead{$\frac{A(\lambda)}{A(V)}$} & \colhead{$\sigma\left(\frac{A(\lambda)}{A(V)}\right)$} & \colhead{$\frac{A(\lambda)}{A(V)}$} & \colhead{$a(\lambda$)} & \colhead{$b(\lambda$)} & \colhead{$\sigma(\lambda)$}}
\startdata
J & 0.269 & 0.018 & 0.268 & 0.258 & -0.636 & 0.037 \\
H & 0.163 & 0.013 & 0.164 & 0.155 & -0.458 & 0.034 \\
K\textsubscript{S} & 0.105 & 0.010 & 0.103 & 0.095 & -0.349 & 0.028 \\
W1 & 0.048 & 0.009 & 0.048 & 0.043 & -0.290 & 0.020 \\
L & 0.045 & 0.008 & 0.046 & 0.041 & -0.274 & 0.020 \\
I1 & 0.044 & 0.009 & 0.045 & 0.040 & -0.270 & 0.021 \\
\hline
0.80 & 0.552 & 0.025 & 0.565 & 0.562 & -0.874 & 0.068 \\
0.85 & 0.523 & 0.023 & 0.509 & 0.504 & -0.629 & 0.056 \\
0.90 & 0.460 & 0.020 & 0.462 & 0.455 & -0.617 & 0.051 \\
0.95 & 0.420 & 0.020 & 0.421 & 0.414 & -0.639 & 0.048 \\
1.00 & 0.381 & 0.020 & 0.386 & 0.377 & -0.647 & 0.045 \\
1.05 & 0.349 & 0.019 & 0.355 & 0.346 & -0.651 & 0.044 \\
1.10 & 0.322 & 0.019 & 0.328 & 0.318 & -0.658 & 0.042 \\
1.15 & 0.304 & 0.019 & 0.304 & 0.294 & -0.668 & 0.039 \\
1.20 & 0.282 & 0.018 & 0.282 & 0.273 & -0.635 & 0.038 \\
1.25 & 0.264 & 0.017 & 0.263 & 0.254 & -0.621 & 0.037 \\
1.30 & 0.248 & 0.017 & 0.246 & 0.236 & -0.632 & 0.036 \\
1.35 & 0.240 & 0.017 & 0.231 & 0.221 & -0.638 & 0.036 \\
1.40 & 0.220 & 0.021 & 0.217 & 0.207 & -0.617 & 0.034 \\
1.45 & 0.210 & 0.019 & 0.204 & 0.195 & -0.576 & 0.033 \\
1.50 & 0.195 & 0.015 & 0.193 & 0.183 & -0.527 & 0.033 \\
1.55 & 0.185 & 0.014 & 0.182 & 0.173 & -0.483 & 0.034 \\
1.60 & 0.172 & 0.013 & 0.173 & 0.163 & -0.461 & 0.034 \\
1.65 & 0.162 & 0.013 & 0.164 & 0.155 & -0.455 & 0.034 \\
1.70 & 0.152 & 0.013 & 0.156 & 0.146 & -0.448 & 0.034 \\
1.75 & 0.145 & 0.013 & 0.148 & 0.139 & -0.439 & 0.034 \\
1.80 & 0.141 & 0.014 & 0.141 & 0.132 & -0.431 & 0.033 \\
1.85 & \nodata & \nodata & 0.135 & 0.126 & -0.424 & 0.032 \\
1.90 & \nodata & \nodata & 0.129 & 0.120 & -0.417 & 0.031 \\
1.95 & 0.125 & 0.012 & 0.123 & 0.115 & -0.408 & 0.030 \\
2.00 & 0.123 & 0.011 & 0.118 & 0.110 & -0.395 & 0.029 \\
2.05 & 0.116 & 0.011 & 0.113 & 0.105 & -0.377 & 0.029 \\
2.10 & 0.110 & 0.011 & 0.108 & 0.100 & -0.359 & 0.029 \\
2.15 & 0.106 & 0.010 & 0.104 & 0.096 & -0.345 & 0.028 \\
2.20 & 0.102 & 0.010 & 0.100 & 0.092 & -0.335 & 0.028 \\
2.25 & 0.097 & 0.010 & 0.096 & 0.089 & -0.332 & 0.027 \\
2.30 & 0.095 & 0.010 & 0.093 & 0.085 & -0.336 & 0.026 \\
2.35 & 0.093 & 0.009 & 0.089 & 0.082 & -0.341 & 0.025 \\
2.40 & 0.090 & 0.010 & 0.086 & 0.079 & -0.347 & 0.025 \\
2.45 & 0.081 & 0.010 & 0.083 & 0.076 & -0.353 & 0.025 \\
2.50 & 0.077 & 0.010 & 0.080 & 0.074 & -0.359 & 0.025 \\
2.55 & \nodata & \nodata & 0.078 & 0.071 & -0.366 & 0.025 \\
2.60 & \nodata & \nodata & 0.075 & 0.069 & -0.372 & 0.025 \\
2.65 & \nodata & \nodata & 0.073 & 0.066 & -0.378 & 0.025 \\
2.70 & \nodata & \nodata & 0.071 & 0.064 & -0.383 & 0.025 \\
2.75 & \nodata & \nodata & 0.068 & 0.062 & -0.386 & 0.024 \\
2.80 & \nodata & \nodata & 0.066 & 0.060 & -0.388 & 0.024 \\
2.85 & \nodata & \nodata & 0.064 & 0.058 & -0.388 & 0.023 \\
2.90 & 0.062 & 0.012 & 0.062 & 0.057 & -0.385 & 0.022 \\
2.95 & 0.058 & 0.012 & 0.061 & 0.055 & -0.379 & 0.021 \\
3.00 & 0.061 & 0.010 & 0.059 & 0.053 & -0.370 & 0.020 \\
3.05 & 0.060 & 0.010 & 0.057 & 0.052 & -0.357 & 0.018 \\
3.10 & 0.059 & 0.009 & 0.056 & 0.050 & -0.342 & 0.017 \\
3.15 & 0.057 & 0.008 & 0.054 & 0.049 & -0.325 & 0.016 \\
3.20 & 0.057 & 0.009 & 0.053 & 0.047 & -0.307 & 0.016 \\
3.25 & 0.055 & 0.008 & 0.051 & 0.046 & -0.290 & 0.016 \\
3.30 & 0.054 & 0.009 & 0.050 & 0.045 & -0.276 & 0.017 \\
3.35 & 0.051 & 0.007 & 0.049 & 0.044 & -0.269 & 0.017 \\
3.40 & 0.050 & 0.007 & 0.048 & 0.043 & -0.266 & 0.018 \\
3.45 & 0.047 & 0.007 & 0.046 & 0.041 & -0.267 & 0.018 \\
3.50 & 0.043 & 0.008 & 0.045 & 0.040 & -0.268 & 0.019 \\
3.55 & 0.039 & 0.008 & 0.044 & 0.039 & -0.269 & 0.020 \\
3.60 & 0.040 & 0.009 & 0.043 & 0.038 & -0.269 & 0.021 \\
3.65 & 0.039 & 0.008 & 0.042 & 0.037 & -0.268 & 0.023 \\
3.70 & 0.039 & 0.008 & 0.041 & 0.037 & -0.266 & 0.024 \\
3.75 & 0.037 & 0.010 & 0.040 & 0.036 & -0.263 & 0.025 \\
3.80 & 0.036 & 0.009 & 0.039 & 0.035 & -0.259 & 0.025 \\
3.85 & 0.033 & 0.009 & 0.038 & 0.034 & -0.253 & 0.025 \\
3.90 & 0.036 & 0.009 & 0.038 & 0.033 & -0.246 & 0.025 \\
3.95 & 0.032 & 0.009 & 0.037 & 0.033 & -0.238 & 0.024 \\
4.00 & 0.032 & 0.008 & 0.036 & 0.032 & -0.229 & 0.022
\enddata
\end{deluxetable}

We also fitted the average curve with a power law and obtained:\footnote{To verify that normalizing to the V-band integrated extinction, $A(V)$, does not affect the shape of the average curve, we also fitted the average curve, normalized to the monochromatic extinction at 1\,\micron, $A(1\,\micron)$, instead of $A(V)$. We find no difference in the power law index of our average extinction curve within the uncertainties.}
\begin{equation}
\label{eq:average}
    \frac{A(\lambda)}{A(V)} = 0.386 \,\lambda^{-1.71}
\end{equation}

\begin{figure*}[ht]
\centering
\includegraphics[width=0.75\textwidth]{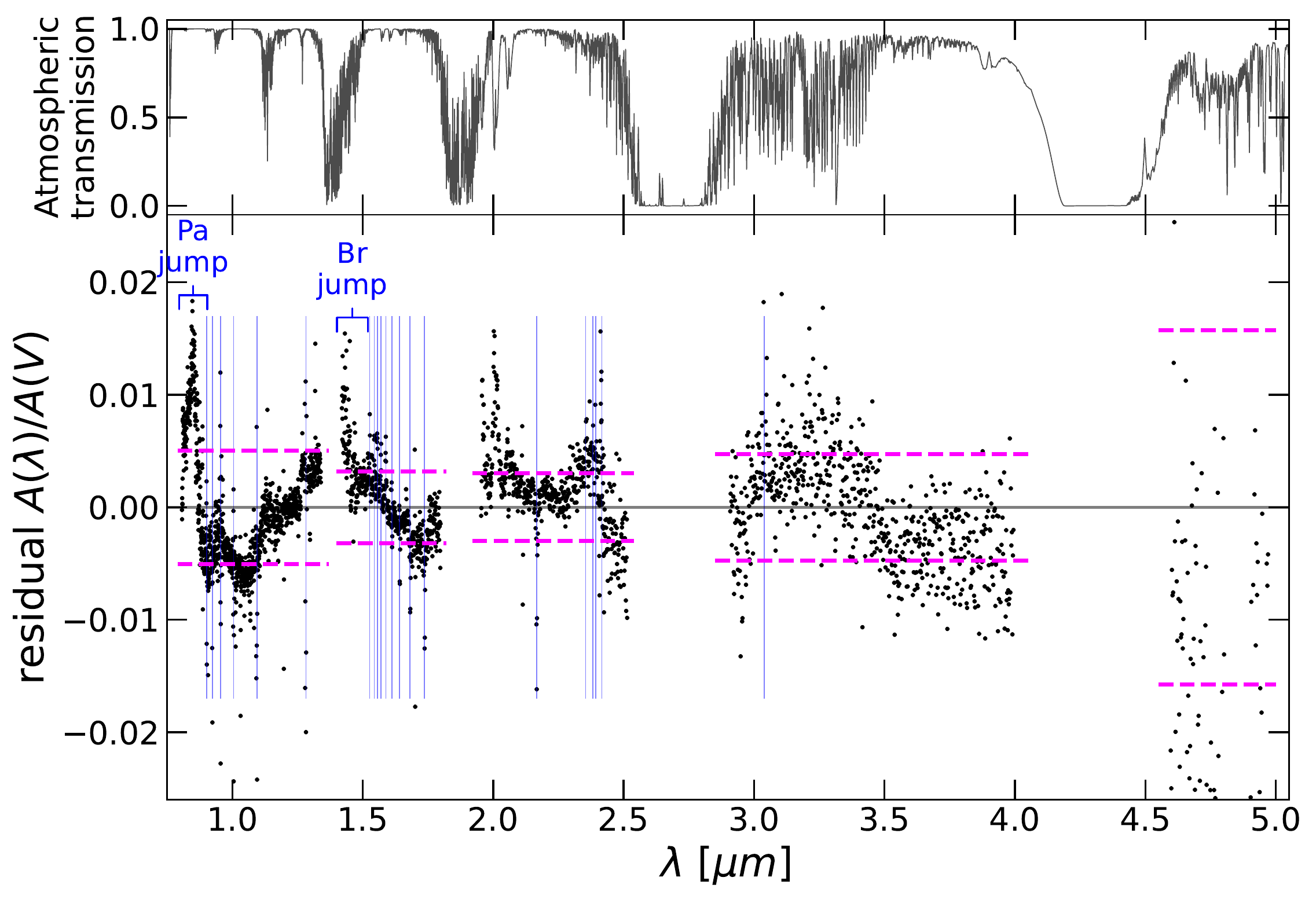}
\caption{Bottom: Residuals \added{(data--fit)} of the average diffuse NIR extinction curve fit (see Fig.~\ref{fig:ave}). The horizontal magenta lines indicate the standard deviations of the residuals in certain wavelength regions: 0.0050 (0.8--1.4\,\micron), 0.0032 (1.4--1.8\,\micron), 0.0030 (1.9--2.5\,\micron), 0.0047 (2.8--4\,\micron) and 0.016 (4.5--5.0\,\micron). The blue vertical lines indicate hydrogen line residuals due to spectral mismatches. The Paschen (Pa) and Brackett (Br) jump mismatches are also indicated in blue. Top: Same atmospheric transmission model as in Fig.~\ref{fig:atmos}. \label{fig:ave_res}}
\end{figure*}

\noindent The fitted parameters are given in Table~\ref{tab:fit_results_diff}, and indicated with a red star in Fig.~\ref{fig:params}. The fitted model is shown in red in Fig.~\ref{fig:ave}, and listed at specific wavelengths and in photometric bands in Table~\ref{tab:RV_dep} (column 4). The fitted model is also available as part of the D22\_MWAvg model in the \texttt{dust\_extinction} python package \citep{dustextinction}. The residuals of the fitting \added{(i.e. the data subtracted by the fit)} are plotted in Fig.~\ref{fig:ave_res}, and are small ($<0.02$) at most wavelengths. The horizontal magenta lines represent the sigma-clipped standard deviation of the residuals in different wavelength regions: 0.0050 (0.8--1.4\,\micron), 0.0032 (1.4--1.8\,\micron), 0.0030 (1.9--2.5\,\micron), 0.0047 (2.8--4\,\micron) and 0.016 (4.5--5.0\,\micron). These values can be considered as the upper limits on the strengths of any features in the average extinction curve. At several specific wavelengths hydrogen lines are visible in the residuals, due to small spectral mismatches between the reddened stars and their comparison star. These are indicated with vertical blue lines. We also indicated peaks in the residuals caused by the Paschen (Pa) and Brackett (Br) jump mismatches. At the top, the same atmospheric transmission model is plotted as in Fig.~\ref{fig:atmos}. Some peaks (e.g. around 2\,\micron) and dips (e.g. just below 3\,\micron) in the residuals can most\deleted{ly} likely be attributed to the telluric absorption.

To verify whether the 3\,\micron\ water ice feature is present in the average diffuse extinction curve, we also fitted it with a combination of a power law and a modified Drude profile with fixed central wavelength, width and asymmetry. The resulting strength of the Drude profile is $B=0.0019\pm0.0007$. This is below the $3\sigma$ detection threshold of \replaced{0.0021\,$A(3\,\micron)/A(V)$}{$A(\text{ice})/A(V)=0.0021$}. However, the strength of the feature is very close to what \cite{2021NatAs...5...78P} have found for the diffuse sightline Cyg~OB2~no.\,12, which adds weight to the possible detection of ice in our diffuse average extinction curve.

\added{
\subsubsection{Comparison to other studies and dust models}
\label{sec:comparison}
Several studies in the past have used a variety of methods to measure NIR extinction curves, and have targeted different Galactic environments. Furthermore, they all had a different way of presenting the extinction measurements, e.g. by giving color excess ratios, reporting absolute or relative extinctions, adopting a different normalization, using various photometric systems, etc. This complicates a comparison between different studies. One thing most of these studies seem to have in common, is that they approximated the NIR extinction curve by a power law, i.e. $A(\lambda)\sim\lambda^{-\alpha}$. Therefore, in this section, we use the power law index $\alpha$ to compare our result with previous measurements. Table~\ref{tab:alphas} compares the fitted power law index for our average diffuse NIR extinction curve to indices found in the literature.

\begin{deluxetable*}{lcccc}
\tablecaption{Comparison of power law indices $\alpha$ from the literature.\label{tab:alphas}}
\tablehead{\colhead{reference} & \colhead{$\alpha$} & \colhead{method used}& \colhead{wavelength range} & \colhead{environment}}
\startdata
this work & $1.71\pm0.01$ & pair method & 0.8--5.0\,\micron\ (spectra) & local diffuse ISM \\
\cite{1989ESASP.290...93D} & $1.75$ & multiple\tablenotemark{a} & 0.9--5\,\micron\ (I, J, H, K, L, M) & Galactic\tablenotemark{a}\\
\cite{1989ApJ...345..245C} & 1.61 & pair method & 0.9--3.3\,\micron\ (I, J, H, K, L) & local ISM\\
\cite{1990ApJ...357..113M} & $1.84\pm0.03$ & pair method & 0.9--4.8\,\micron\ (I, J, H, K, L, M) & local diffuse ISM \\
\cite{2006ApJ...638..839N} & $1.99\pm0.02$ & RC CMD & 1.2--2.2\,\micron\ (J, H, K\textsubscript{S}) & Galactic center\\
\cite{2009ApJ...696.1407N} & 2.0 & RC+RGB CMD & 1.2--2.2\,\micron\ (J, H, K\textsubscript{S}) & Galactic center\\
\cite{2009MNRAS.400..731S} & $2.14^{+0.04}_{-0.05}$ & RC CCD & 1.2--2.2\,\micron\ (J, H, K\textsubscript{S}) & Galactic plane\\
\cite{2011ApJ...737...73F} & $2.13 \pm 0.08$ & nebular H lines & 1.282--2.166\,\micron & Galactic center\\
 & $1.76 \pm 0.39$ & & 2.166--2.758\,\micron\\
 & $2.11 \pm 0.06$ & & 1.282--2.758\,\micron\\
\cite{2017ApJ...849L..13A} & $2.47\pm0.11$ & RC CMD & 0.88--2.15\,\micron\ (Z, Y, J, H, K\textsubscript{S}) & Galactic center\\
\cite{2018AA...610A..83N} & $2.30\pm0.08$ & RC\tablenotemark{b} & 1.27--2.16\,\micron\ (J, H, K\textsubscript{S}) & Galactic center\\
\cite{2019AA...630L...3N} & $2.43\pm0.03$ & RC\tablenotemark{b} & 1.27--1.65\,\micron\ (J, H) & Galactic center\\
 & $2.23\pm0.03$ & & 1.65--2.16\,\micron\ (H, K\textsubscript{S}) \\
 & $2.32\pm0.09$ & & 1.27--2.16\,\micron\ (J, H, K\textsubscript{S})\\
\cite{2020MNRAS.496.4951M} & 2.27 & RC CCD & 1.2--2.2\,\micron\ (J, H, K\textsubscript{S}) & Galactic plane \\
\enddata
\tablenotetext{a}{This paper combined several extinction measurements from the literature, obtained with different methods and for different Galactic environments.}
\tablenotetext{b}{This work used different methods based on RC stars. We report their average $\alpha$ value.}
\end{deluxetable*}

Our value for $\alpha$ is in reasonable agreement with that from \cite{1989ESASP.290...93D}, and \replaced{not very deviant from}{between} the results of \cite{1989ApJ...345..245C} and \cite{1990ApJ...357..113M}. Later measurements toward the GC have a larger power law index compared to our result. It has to be noted that those measurements probe highly extinguished ($A(V)\gtrsim8$) regions in the inner Galactic disk and bulge, at about 8\,kpc distance. On the contrary, we measure extinction in the local (closer than 3\,kpc, see distances in Table~\ref{tab:sample}), low-extinction ($A(V)<3.9$), diffuse ISM. The very long sightlines toward the distant GC have a higher probability of encountering molecular clouds along the line-of-sight compared to our local measurements. Those sightlines thus likely contain a mix of diffuse and dense material, which can have an effect on the shape of the extinction curve. The presence of dense material towards the GC is confirmed by the detection of ice features \citep[see e.g.,][]{2011ApJ...737...73F}, which are likely only (clearly) visible in sightlines through molecular clouds and not in purely diffuse extinction regions \citep[][and see our discussion in Section \ref{sec:features}]{1997ApJ...490..729W}. On the other hand, the detection of the strong aliphatic hydrocarbon feature at 3.4\,\micron\ shows that those sightlines also contain diffuse dust \citep{2011ApJ...737...73F}. We are confident that our average curve probes purely diffuse extinction, given that no strong ice feature has been detected and that our average UV extinction curve is close to diffuse curves from the literature (see Fig.~\ref{fig:ave}).
\cite{2011ApJ...737...73F} also suggested that it is likely that there is a transition of a mostly flatter NIR extinction curve in the solar neighborhood to a steeper one in most parts of the Galactic disk.

We want to emphasize that we find a quite large range of indices within our sample of diffuse sightlines (between 1.36 and 2.20, see Table \ref{tab:fit_results_diff}), showing that there are real sightline-to-sightline variations in the shape of the NIR extinction curve within the local diffuse ISM. \cite{2006ApJ...638..839N, 2009ApJ...696.1407N}, \cite{2011ApJ...737...73F}, and \cite{2017ApJ...849L..13A} also found that the NIR extinction curve changes from one sightline to another. As a consequence, the shape of any average extinction curve will depend on the sightlines that were used to measure that average, and it is not surprising that different studies, using different samples, find different average curves. As can be seen from Fig.~\ref{fig:params}, and as briefly discussed in Section~\ref{sec:trends}, the variation in $\alpha$ seems to be anti-correlated with $R(V)$. A more detailed discussion on the $R(V)$-dependence of the NIR extinction curve is given in Section \ref{sec:RV_dep}.

As mentioned in the introduction, some studies toward the GC (and Galactic plane) reported a dependence of the power law index on the wavelength region. For example, \cite{2005ApJ...619..931I} and \cite{2009ApJ...696.1407N} saw a flattening of the extinction curve at wavelengths beyond $\sim3$\,\micron. \cite{2019AA...630L...3N} measured a smaller index $\alpha$ in HK\textsubscript{S} than in JH (see Table \ref{tab:alphas}). \cite{2011ApJ...737...73F} obtained a flatter extinction curve beyond $\sim3.7$\,\micron. They also found a smaller $\alpha$ between 2.2 and 2.8\,\micron\ than between 1.2 and 2.2\,\micron\ (see Table \ref{tab:alphas}), but stated that this change in slope is not significant.
We demonstrate that our local average diffuse extinction curve between 0.8 and 4\,\micron\ can be represented by a single power law, since there are no long-range trends visible in the residuals in Fig.~\ref{fig:ave_res}. We argue that it is possible that the mix of diffuse and dense material in sightlines toward the GC and in the Galactic plane imposes the need for multiple power law indices across the NIR.}

\begin{figure*}[ht]
\centering
\includegraphics[width=0.6\textwidth]{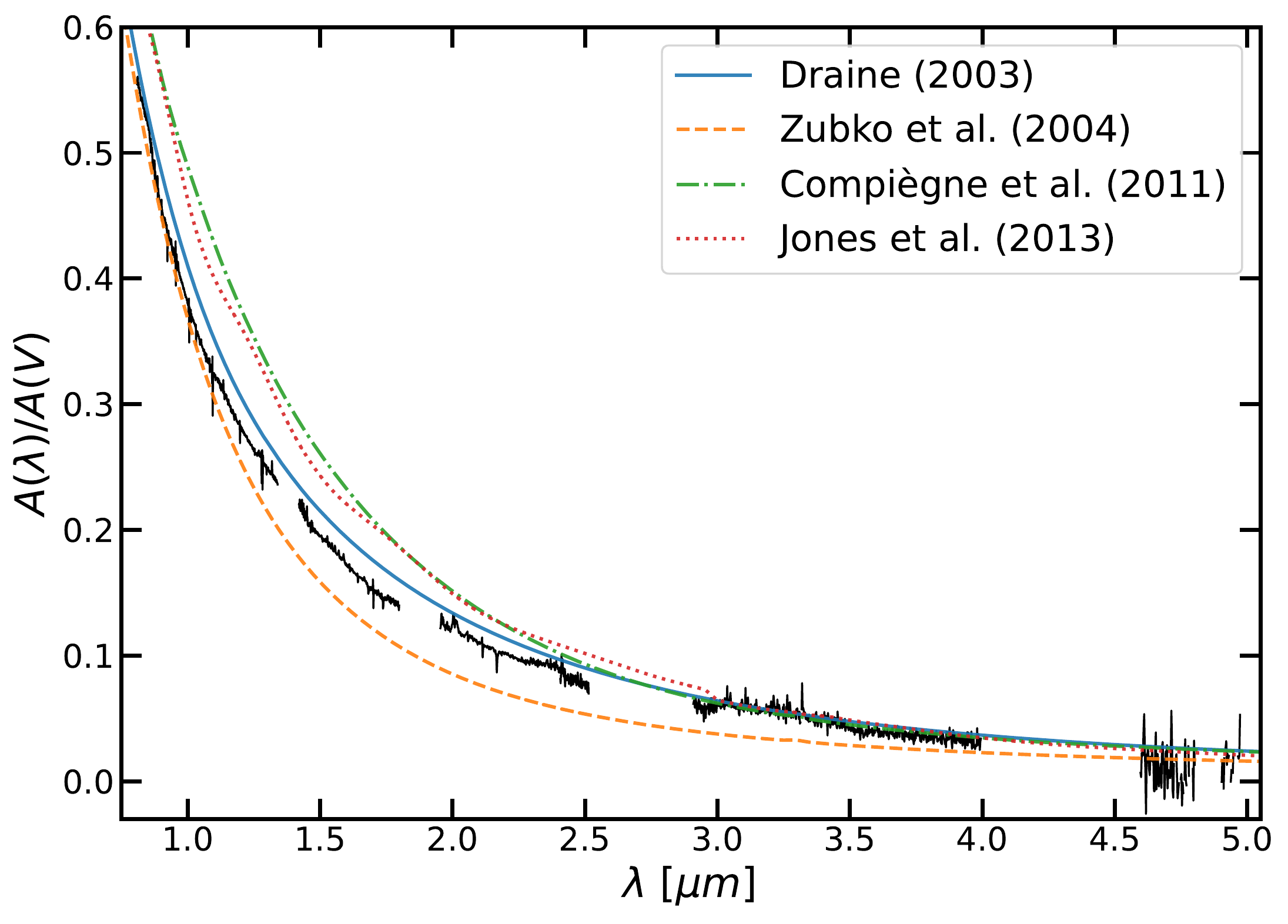}
\caption{Average diffuse NIR extinction curve, compared to the \cite{2003ARA&A..41..241D, 2003ApJ...598.1017D}, \cite{2004ApJS..152..211Z}, \cite{2011A&A...525A.103C} and \cite{2013A&A...558A..62J} diffuse ISM dust grain models with $R(V) = 3.1$.
The \cite{2001ApJ...548..296W} model is not shown in this plot as it is very similar to the \cite{2003ARA&A..41..241D, 2003ApJ...598.1017D} model. \label{fig:ave_mod}}
\end{figure*}

In Fig.~\ref{fig:ave_mod}, we compare our \added{average} diffuse \deleted{average} NIR extinction curve to the \cite{2003ARA&A..41..241D, 2003ApJ...598.1017D}, \cite{2004ApJS..152..211Z}, \cite{2011A&A...525A.103C} and \cite{2013A&A...558A..62J} diffuse ISM dust grain models. The \cite{2004ApJS..152..211Z} model seems to agree with \replaced{the}{our} average curve below 1\,\micron, while the other models correspond better beyond 3\,\micron.

\subsection{The R(V)-dependent NIR extinction curve}
\label{sec:RV_dep}
Although the average extinction curve described in the previous section can be very useful to get an idea of the overall shape of the diffuse NIR extinction curve, we found clear variations around this average for different sightlines, as can be seen from the fitting results in Table~\ref{tab:fit_results_diff} and in Fig.~\ref{fig:params}. \added{Because of these strong sightline-to-sightline variations, the shape of any average extinction curve will depend on the sample of sightlines used to measure that average, so one should be cautious when using an average extinction curve.} Interestingly, several previous studies have shown that these variations are largely dependent upon a single parameter, often chosen to be $R(V)$. \cite{1989ApJ...345..245C} presented a linear relationship between the extinction curve and $1/R(V)$, based on photometric extinction measurements between 0.1 and 3.4\,\micron. For the NIR region, they found that the slopes and intercepts of this linear relationship can be described by a power law. More recently, \cite{2019ApJ...886..108F} presented a linear relationship between the extinction curve and $R(V)$ based on a combination of spectral UV and optical extinction measurements and 2MASS JHK\textsubscript{S} measurements.

\begin{figure}[t]
\centering
\includegraphics[width=0.9\columnwidth]{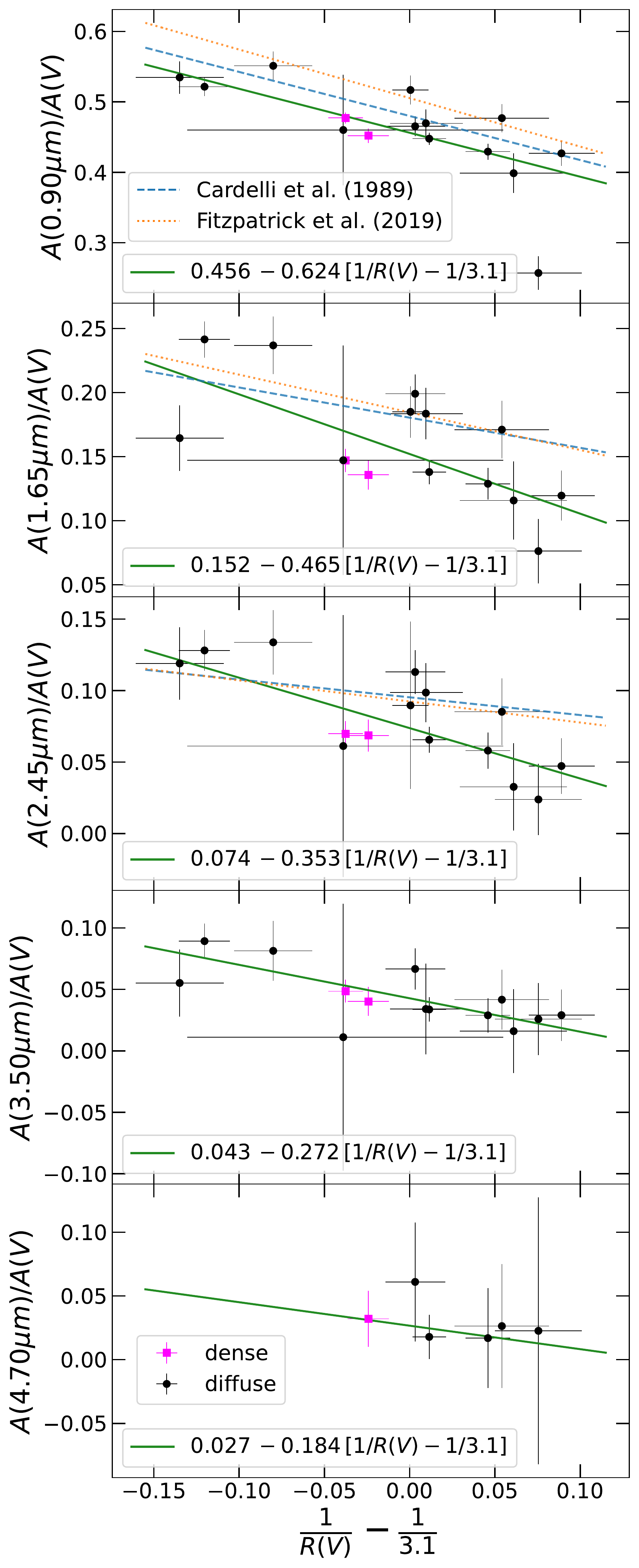}
\caption{Illustration of the $R(V)$-dependence of the extinction curve at a handful of wavelengths. \added{The magenta squares indicate the dense sightlines and the black circles the diffuse sightlines.} The linear fits are shown as green \deleted{dashed} lines. \added{For comparison, the $R(V)$-dependent relationships from \cite{1989ApJ...345..245C} and \cite{2019ApJ...886..108F} are added as blue dashed and orange dotted lines, respectively.} \label{fig:RV_dep}}
\end{figure}

In this section, we investigate our sample of observed NIR extinction curves as a function of \replaced{$R(V)$}{$1/R(V)$}.\footnote{One could also use $R(V)$, but we found that $1/R(V)=E(B-V)/A(V)$ more naturally results in a linear relationship with the normalized extinction $A(\lambda)/A(V)$, given that both quantities have the same denominator.} Fig.~\ref{fig:RV_dep} shows $A(\lambda)/A(V)$ vs.\ \replaced{$R(V)-3.1$}{$1/R(V)-1/3.1$} at 5 different wavelengths (chosen in regions where the atmospheric absorption is not significantly affecting the extinction measurement). We indeed see a linear trend at all wavelengths. It has to be noted, however, that at the longest wavelength (4.7\,\micron, bottom plot), the range in $R(V)$ values is limited due to the limited number of sightlines with data at wavelengths beyond 4\,\micron. 

To quantify this correlation, \added{using the \texttt{LinearLSQFitter} from Astropy,} we fitted a linear function \added{(\texttt{Linear1D} Astropy model)} at every wavelength in the SpeX range where at least 5 sightlines have data ($\sim$2800 data points between 0.8 and 5.1\,\micron):

\begin{equation}
    \frac{A(\lambda)}{A(V)} = a(\lambda) + b(\lambda) \,\left[\frac{1}{R(V)}-\frac{1}{3.1}\right] 
\end{equation}

\begin{figure*}[ht]
\centering
\includegraphics[width=0.7\textwidth]{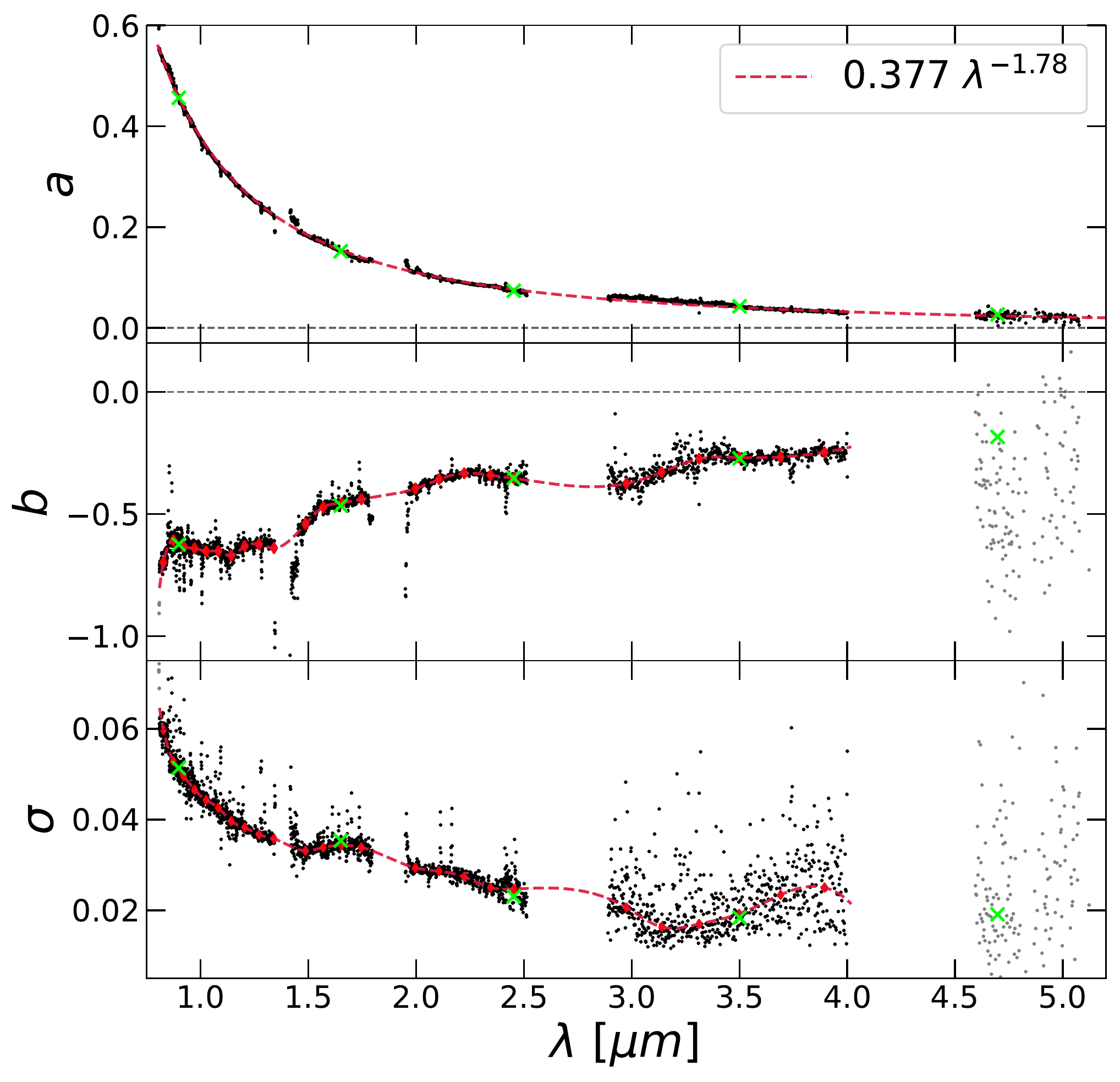}
\caption{Intercepts $a$ (top), slopes $b$ (middle) and standard deviations about the fit $\sigma$ (bottom) of the linear \replaced{$R(V)$}{$1/R(V)$}-dependence, as a function of wavelength. The green crosses indicate the wavelengths plotted in Fig.~\ref{fig:RV_dep}. The red dashed line in the top panel shows the fitted power law to the intercepts. The red diamonds in the middle and bottom plot are the anchor points used in the spline interpolation, while the red dashed line represents the cubic spline interpolation. In these panels, data points above 4\,\micron\ are plotted in gray and were excluded from the interpolation because they are very noisy. \label{fig:slope_inter}}
\end{figure*}

\noindent By choosing \replaced{$R(V)-3.1$}{$1/R(V)-1/3.1$} as the abscissa, the intercepts $a(\lambda)$ correspond to the average $R(V)=3.1$ Milky Way extinction curve. The slopes $b(\lambda)$ illustrate the $R(V)$-dependence. \added{The fits are shown as green lines in Fig.~\ref{fig:RV_dep}. We also added the $R(V)$-dependent relationships from \cite{1989ApJ...345..245C} and \cite{2019ApJ...886..108F} (where available) for comparison. At $\lambda=0.9\,\micron$ these have a similar slope to our fitted line, while at longer wavelengths our fit is steeper than the literature lines, which means that we find a stronger dependence on $R(V)$.} We also calculated the standard deviation of the individual sightlines from the fitted relationship, which reflects both measurement uncertainties and real deviations from the $R(V)$-dependent relationship. Fig.~\ref{fig:slope_inter} shows the obtained intercepts, slopes and standard deviations as a function of wavelength. It is clear that the $R(V)$-dependence of the extinction slowly decreases towards longer wavelengths, but it does not entirely disappear within our wavelength range.

We want to note here that the fact that $A(\lambda)/A(V)$ and $1/R(V)=E(B-V)/A(V)$ have a common factor, namely $A(V)$, could possibly lead to an artificial correlation between both quantities and between their uncertainties. However, we found that the correlation coefficients between both quantities, due to their common factor, are small, ranging from -0.02 to 0.12, with a median value over all sightlines and all NIR wavelengths of 0.02. More importantly, the correlation coefficients are \textit{positive} for most wavelengths and most sightlines. In contrast, the observed relationship between both quantities is a \textit{negative} linear trend (i.e. an anti-correlation) as $b(\lambda)<0$ for all NIR wavelengths (see Figs.~\ref{fig:RV_dep} and \ref{fig:slope_inter}), and could thus not be caused by the (small) positive artificial correlation due to their common factor $A(V)$.\footnote{Note that this is different at UV wavelengths, see e.g. \cite{2007ApJ...663..320F}.}

In order to provide a smooth representation of the $R(V)$-dependence for use in future extinction studies, \added{using the \texttt{LevMarLSQFitter} from Astropy,} we fitted the intercepts with a power law \added{(\texttt{PowerLaw1D} Astropy model)} and found:
\begin{equation}
    a(\lambda) = 0.377 \,\lambda^{-1.78}
\end{equation}

\noindent As mentioned above, the intercepts correspond to the average $R(V)=3.1$ Milky Way extinction curve. Hence, it is not surprising that they are well fitted with a power law function, similar to the average curve described in Section \ref{sec:ave_fit}. For the slopes and standard deviations, no straightforward functional form could be fitted. Instead, we interpolated the data with cubic splines using \texttt{scipy} \citep{2020SciPy-NMeth}. We first binned the data into 25 equally-sized wavelength bins (i.e. every bin has approximately 110 data points), and calculated the median wavelength, median slope, and median standard deviation in every bin. These median values were then used as anchor points for the spline interpolation. The data points above 4\,\micron\ (plotted in gray in Fig.~\ref{fig:slope_inter}) are very noisy and were not included in the interpolation. The fitted intercepts, and interpolated slopes and standard deviations are shown as red dashed lines in Fig.~\ref{fig:slope_inter}, and are tabulated in Table~\ref{tab:RV_dep} (columns 5--7) for wavelengths between 0.8 and 4\,\micron, and in common IR photometric bands (2MASS J, H, K\textsubscript{S}, WISE~1, L, and IRAC 1), obtained by integrating the fitted/interpolated curves over the response curves of the bands. It has to be noted here that the broad ``wiggles" in the slopes are most likely a result of the telluric absorption. Our $R(V)$-dependent model is also available as the D22 model in the \texttt{dust\_extinction} python package \citep{dustextinction}.

\begin{figure}[ht]
\centering
\includegraphics[width=\columnwidth]{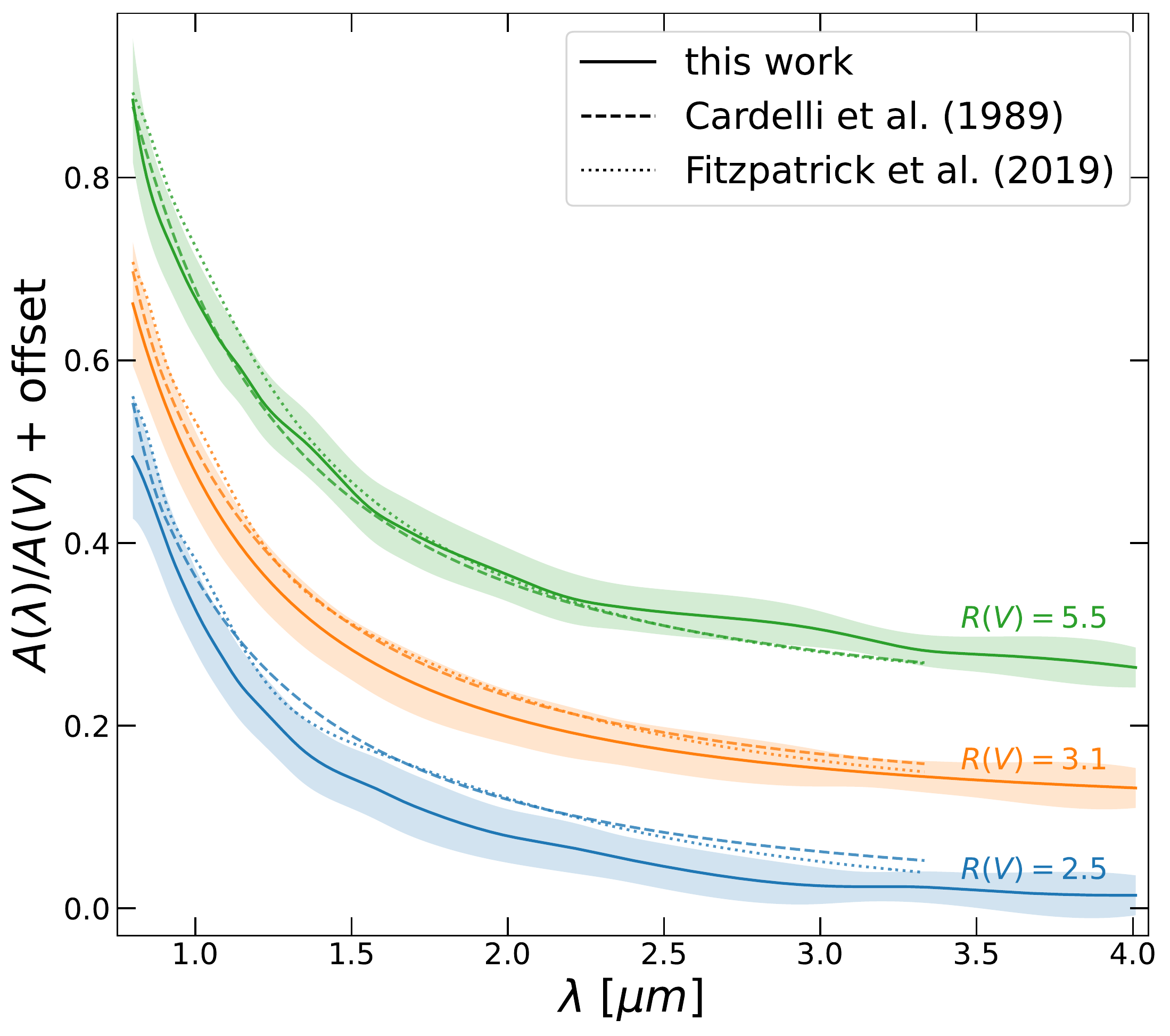}
\caption{$R(V)$-dependent extinction curves for different $R(V)$ values: 2.5 in blue, 3.1 in orange and 5.5 in green, in comparison with the $R(V)$-dependent curves from \cite{1989ApJ...345..245C} in dashed lines and \cite{2019ApJ...886..108F} in dotted lines. The shaded regions represent the standard deviation around the curve.
 \label{fig:RV_lit}}
\end{figure}

Fig.~\ref{fig:RV_lit} shows the derived $R(V)$-dependent NIR extinction curve for 3 values of $R(V)$ (2.5, 3.1 and 5.5). The shaded regions represent the standard deviation about the curve (last column of Table~\ref{tab:RV_dep}). For comparison, we added the $R(V)$-dependent curves from \cite{1989ApJ...345..245C} and \cite{2019ApJ...886..108F}, which mostly fall within the shaded regions \added{for $R(V)=3.1$ and $R(V)=5.5$. However, for $R(V)=2.5$ the literature curves deviate slightly from our curve for wavelengths $\gtrsim1\,\micron$, as was also clear from Fig.~\ref{fig:RV_dep}}.

It has been suggested in the past that the $R(V)$-dependence of NIR extinction curves is not real, but simply a result of the normalization of the extinction curves to $A(V)$, and disappears when normalizing at longer wavelengths \citep[see e.g.,][]{1989ApJ...345..245C}. We tested this by normalizing the curves to the extinction at 1\,\micron, and plotting the extinction as a function of \replaced{$R(V)-3.1$}{$1/R(V)-1/3.1$} as before. We still found a strong linear correlation between extinction and \replaced{$R(V)$}{$1/R(V)$}.

We conclude that the $R(V)$-dependent extinction curve can reduce uncertainties when applying extinction corrections to sightlines with known $R(V)$ values, because it takes into account sightline variations which are correlated with $R(V)$. Real variations around the $R(V)$-dependent curve are still present, but smaller than the variations around the average diffuse extinction curve derived in the previous section. \added{We thus recommend using our $R(V)$-dependent extinction curve when possible.}

\added{This work completes a series of spectroscopic extinction curve studies from the far-UV to the MIR \citep{2009ApJ...705.1320G,2019ApJ...886..108F,2021ApJ...916...33G}. In future work, we will combine these results to create an $R(V)$-dependent Milky Way extinction curve from 912\,\AA\ to 30\,\micron\ (Gordon et al., in prep.). One focus of this future work will be to investigate the impact of underlying correlations between $A(\lambda)/A(V)$ and $1/R(V)$ (and between their uncertainties) on the $R(V)$ relationship. Preliminary results show that such correlations mainly impact wavelengths shorter than 1\,\micron, confirming that the observed relationship in this work (in the NIR) is not significantly affected by underlying correlations.}

\subsection{Extinction features}
\label{sec:features_lab}
Several NIR-MIR observations towards dense interstellar clouds have revealed features in the extinction curves that are caused by ices. \cite{2015ARA&A..53..541B} provide a detailed review on the detected ice features in infrared spectra. The strongest ice feature has a peak wavelength around 3\,\micron, and is primarily caused by the \mbox{O-H} stretching modes of bulk H\textsubscript{2}O ice. While a detailed study of this feature is beyond the scope of this work, we briefly discuss its characteristics in our sample.

The 3\,\micron\ ice feature was only clearly detected in the two dense sightlines, \object{HD283809} and \object{HD029647}, (and very tentatively in \object{HD183143} and \object{HD229238}). Both \object{HD283809} and \object{HD029647} lie toward the Taurus Dark Cloud region, close to the dense clump TMC--1, and the presence of ice features in their spectra has been previously reported \citep[e.g.,][]{1988MNRAS.233..321W, 1993MNRAS.263..749S}. We show in Section~\ref{sec:features} that the feature can be fitted reasonably well with a modified Drude profile. Fig.~\ref{fig:features} shows that the feature has a long-wavelength wing. As explained by \cite{2015ARA&A..53..541B}, large grains could be responsible for part but not all of the wing. The remainder of the absorption may be attributed to the \mbox{O-H} stretching mode in ammonia hydrates \citep{1982ApJ...260..141K, 2001A&A...365..144D}.

Structure near 3.4\,\micron\ might arise from the \mbox{C-H} stretching mode in hydrocarbons \citep{2002ApJS..138...75P}, which have been detected along diffuse sightlines towards a few background stars \citep{1994ApJ...437..683P}. If so, it is possible that our approach to fit the ice feature masks this weaker hydrocarbon feature which could arise from the diffuse dust along these sightlines.

It has to be noted that the telluric absorption (see Fig.~\ref{fig:atmos}) severely complicates the measurement of any features in the NIR extinction curve. It is, for example, very difficult to constrain the underlying continuum extinction and the shape of the feature, because of the gap between 2.5 and 2.88\,\micron. Also, the data between 2.88 and 3.5\,\micron\ are very noisy. Fortunately, \added{our approved cycle 1 GO program 2459\footnote{\url{https://www.stsci.edu/jwst/phase2-public/2459.pdf}} and other programs} on the James Webb Space Telescope will provide continuous, well calibrated, high signal-to-noise spectra of dust extinguished Milky Way OB stars, which will enable a much more detailed study of dust extinction features in the NIR (and MIR).

\section{Conclusions}
\label{sec:conclusions}
In this work, we presented the first comprehensive spectroscopic study of NIR extinction curves in the Milky Way. We obtained spectra with the IRTF/SpeX spectrograph between 0.8 and 5.5\,\micron\ for a sample of 25 reddened and 15 comparison stars. We were able to measure extinction curves for 15 sightlines using the pair method. This sample spans $A(V)$ values from 0.78 to 5.65 and $R(V)$ values from \replaced{2.33}{2.43} to 5.33. The main conclusions of this work are:
\begin{itemize}
    \item The NIR extinction curves in our sample are well fitted by a single power law over the entire wavelength range between 0.8 and 5.2\,\micron, with indices and amplitudes \added{strongly} differing from sightline to sightline. 
    \item \replaced{The}{Our} average \added{local} diffuse NIR \added{Milky Way} extinction curve can be represented by a single power law with index $\alpha=1.7$ over the entire wavelength range between 0.8 and 5\,\micron, and agrees well with the average curve from \cite{1990ApJ...357..113M}.
    \item \added{The shape of any average extinction curve depends on the sample of sightlines used to measure the average, and one should thus be cautious when using an average extinction curve.}
    \item We find that the \added{normalized} extinction curves \added{$A(\lambda)/A(V)$} in our sample linearly depend on the \replaced{total-to-selective extinction $R(V)$}{selective-to-total extinction $1/R(V)$}, with the strength of the dependence decreasing with increasing wavelength.
    \item Two sightlines in our sample (\object{HD283809} and \object{HD029647}) show \replaced{an}{a strong} ice extinction feature around 3\,\micron, which can be approximated by a modified Drude profile. These sightlines most likely contain dense material\deleted{, and our fitting approach might mask a weak 3.4\,\micron\ hydrocarbon feature known to emanate from diffuse interstellar dust}.
    \item We tentatively detect (with slightly over 3$\sigma$ significance) the 3\,\micron\ ice feature in the extinction curves for \object{HD183143} and \object{HD229238}. These sightlines have the highest $A(V)$ values of the diffuse sample.
    \item We do not detect the 3\,\micron\ ice feature in the average diffuse extinction curve, with a $3\sigma$ limit of \replaced{0.0021\,$A(3\,\micron)/A(V)$}{$A(\text{ice})/A(V)=0.0021$}.
    \item The telluric atmospheric absorption complicates the characterization of any extinction features, as well as the continuum extinction above 4\,\micron. Planned cycle 1 observations with the James Webb Space Telescope will provide significant improvements to our understanding of the NIR dust extinction.
\end{itemize}

\added{The SpeX spectra and measured extinction curves are electronically \dataset[available]{\doi{10.5281/zenodo.5802469}} \citep{decleir_marjorie_2022_5802469}.
The code used for the analysis and plots is available on GitHub \citep{marjorie_decleir_2022_6330037}.}

\acknowledgements
We thank the referee for their feedback, which improved this paper.
This research has made use of the SIMBAD database, operated at CDS, Strasbourg, France.
This research has made use of the NASA/IPAC Infrared Science Archive, which is funded by the National Aeronautics and Space Administration and operated by the California Institute of Technology.
This publication makes use of data products from the Two Micron All Sky Survey, which is a joint project of the University of Massachusetts and the Infrared Processing and Analysis Center/California Institute of Technology, funded by the National Aeronautics and Space Administration and the National Science Foundation.
\added{This publication makes use of data products from the Wide-field Infrared Survey Explorer, which is a joint project of the University of California, Los Angeles, and the Jet Propulsion Laboratory/California Institute of Technology, funded by the National Aeronautics and Space Administration.}
This research has made use of the SVO Filter Profile Service (\url{http://svo2.cab.inta-csic.es/theory/fps/}) supported from the Spanish MINECO through grant AYA2017-84089.

\facilities{IRTF (SpeX), IRSA, MAST}

\software{Spextool \citep{2004PASP..116..362C}, Astropy \citep{2013A&A...558A..33A}, measure\_extinction \citep{measureextinction}, dust\_extinction \citep{dustextinction}, spex\_nir\_extinction \citep{marjorie_decleir_2022_6330037}}

\bibliography{references}{}

\begin{thebibliography}{}
\expandafter\ifx\csname natexlab\endcsname\relax\def\natexlab#1{#1}\fi
\providecommand{\url}[1]{\href{#1}{#1}}
\providecommand{\dodoi}[1]{doi:~\href{http://doi.org/#1}{\nolinkurl{#1}}}
\providecommand{\doeprint}[1]{\href{http://ascl.net/#1}{\nolinkurl{http://ascl.net/#1}}}
\providecommand{\doarXiv}[1]{\href{https://arxiv.org/abs/#1}{\nolinkurl{https://arxiv.org/abs/#1}}}

\bibitem[{{Abt} \& {Golson}(1962)}]{1962ApJ...136...35A}
{Abt}, H.~A., \& {Golson}, J.~C. 1962, \apj, 136, 35, \dodoi{10.1086/147349}

\bibitem[{{Alonso-Garc{\'\i}a} {et~al.}(2017){Alonso-Garc{\'\i}a}, {Minniti},
  {Catelan}, {Contreras Ramos}, {Gonzalez}, {Hempel}, {Lucas}, {Saito},
  {Valenti}, \& {Zoccali}}]{2017ApJ...849L..13A}
{Alonso-Garc{\'\i}a}, J., {Minniti}, D., {Catelan}, M., {et~al.} 2017, \apjl,
  849, L13, \dodoi{10.3847/2041-8213/aa92c3}

\bibitem[{{Astropy Collaboration} {et~al.}(2013){Astropy Collaboration},
  {Robitaille}, {Tollerud}, {Greenfield}, {Droettboom}, {Bray}, {Aldcroft},
  {Davis}, {Ginsburg}, {Price-Whelan}, {Kerzendorf}, {Conley}, {Crighton},
  {Barbary}, {Muna}, {Ferguson}, {Grollier}, {Parikh}, {Nair}, {Unther},
  {Deil}, {Woillez}, {Conseil}, {Kramer}, {Turner}, {Singer}, {Fox}, {Weaver},
  {Zabalza}, {Edwards}, {Azalee Bostroem}, {Burke}, {Casey}, {Crawford},
  {Dencheva}, {Ely}, {Jenness}, {Labrie}, {Lim}, {Pierfederici}, {Pontzen},
  {Ptak}, {Refsdal}, {Servillat}, \& {Streicher}}]{2013A&A...558A..33A}
{Astropy Collaboration}, {Robitaille}, T.~P., {Tollerud}, E.~J., {et~al.} 2013,
  \aap, 558, A33, \dodoi{10.1051/0004-6361/201322068}

\bibitem[{{Boogert} {et~al.}(2015){Boogert}, {Gerakines}, \&
  {Whittet}}]{2015ARA&A..53..541B}
{Boogert}, A.~C.~A., {Gerakines}, P.~A., \& {Whittet}, D. C.~B. 2015, \araa,
  53, 541, \dodoi{10.1146/annurev-astro-082214-122348}

\bibitem[{{Boogert} {et~al.}(2011){Boogert}, {Huard}, {Cook}, {Chiar}, {Knez},
  {Decin}, {Blake}, {Tielens}, \& {van Dishoeck}}]{2011ApJ...729...92B}
{Boogert}, A.~C.~A., {Huard}, T.~L., {Cook}, A.~M., {et~al.} 2011, \apj, 729,
  92, \dodoi{10.1088/0004-637X/729/2/92}

\bibitem[{{Borgman}(1960)}]{1960BAN....15..255B}
{Borgman}, J. 1960, \bain, 15, 255

\bibitem[{{Cardelli} {et~al.}(1989){Cardelli}, {Clayton}, \&
  {Mathis}}]{1989ApJ...345..245C}
{Cardelli}, J.~A., {Clayton}, G.~C., \& {Mathis}, J.~S. 1989, \apj, 345, 245,
  \dodoi{10.1086/167900}

\bibitem[{{Chiar} \& {Tielens}(2006)}]{2006ApJ...637..774C}
{Chiar}, J.~E., \& {Tielens}, A.~G.~G.~M. 2006, \apj, 637, 774,
  \dodoi{10.1086/498406}

\bibitem[{{Clayton} {et~al.}(2015){Clayton}, {Gordon}, {Bianchi}, {Massa},
  {Fitzpatrick}, {Bohlin}, \& {Wolff}}]{2015ApJ...815...14C}
{Clayton}, G.~C., {Gordon}, K.~D., {Bianchi}, L.~C., {et~al.} 2015, \apj, 815,
  14, \dodoi{10.1088/0004-637X/815/1/14}

\bibitem[{{Clayton} {et~al.}(2003){Clayton}, {Gordon}, {Salama}, {Allamandola},
  {Martin}, {Snow}, {Whittet}, {Witt}, \& {Wolff}}]{2003ApJ...592..947C}
{Clayton}, G.~C., {Gordon}, K.~D., {Salama}, F., {et~al.} 2003, \apj, 592, 947,
  \dodoi{10.1086/375771}

\bibitem[{{Compi{\`e}gne} {et~al.}(2011){Compi{\`e}gne}, {Verstraete}, {Jones},
  {Bernard}, {Boulanger}, {Flagey}, {Le Bourlot}, {Paradis}, \&
  {Ysard}}]{2011A&A...525A.103C}
{Compi{\`e}gne}, M., {Verstraete}, L., {Jones}, A., {et~al.} 2011, \aap, 525,
  A103, \dodoi{10.1051/0004-6361/201015292}

\bibitem[{{Crawford} {et~al.}(1971){Crawford}, {Barnes}, \&
  {Golson}}]{1971AJ.....76.1058C}
{Crawford}, D.~L., {Barnes}, J.~V., \& {Golson}, J.~C. 1971, \aj, 76, 1058,
  \dodoi{10.1086/111220}

\bibitem[{{Cushing} {et~al.}(2004){Cushing}, {Vacca}, \&
  {Rayner}}]{2004PASP..116..362C}
{Cushing}, M.~C., {Vacca}, W.~D., \& {Rayner}, J.~T. 2004, \pasp, 116, 362,
  \dodoi{10.1086/382907}

\bibitem[{{Dartois} \& {d'Hendecourt}(2001)}]{2001A&A...365..144D}
{Dartois}, E., \& {d'Hendecourt}, L. 2001, \aap, 365, 144,
  \dodoi{10.1051/0004-6361:20000174}

\bibitem[{Decleir(2022{\natexlab{a}})}]{decleir_marjorie_2022_5802469}
Decleir, M. 2022{\natexlab{a}}, SpeX NIR spectra and extinction curves, 2.0,
  Zenodo, \dodoi{10.5281/zenodo.5802469}

\bibitem[{Decleir(2022{\natexlab{b}})}]{marjorie_decleir_2022_6330037}
---. 2022{\natexlab{b}}, {mdecleir/spex\_nir\_extinction: Release related to
  final SpeX NIR extinction paper}, v1.1.0,  Zenodo,
  \dodoi{10.5281/zenodo.6330037}

\bibitem[{{Draine}(1989)}]{1989ESASP.290...93D}
{Draine}, B.~T. 1989, in Infrared Spectroscopy in Astronomy, ed.
  E.~{B{\"o}hm-Vitense}, 93

\bibitem[{{Draine}(2003{\natexlab{a}})}]{2003ARA&A..41..241D}
{Draine}, B.~T. 2003{\natexlab{a}}, \araa, 41, 241,
  \dodoi{10.1146/annurev.astro.41.011802.094840}

\bibitem[{{Draine}(2003{\natexlab{b}})}]{2003ApJ...598.1017D}
---. 2003{\natexlab{b}}, \apj, 598, 1017, \dodoi{10.1086/379118}

\bibitem[{{Fernie}(1983)}]{1983ApJS...52....7F}
{Fernie}, J.~D. 1983, \apjs, 52, 7, \dodoi{10.1086/190856}

\bibitem[{{Fitzgerald}(1970)}]{1970A&A.....4..234F}
{Fitzgerald}, M.~P. 1970, \aap, 4, 234

\bibitem[{{Fitzpatrick}(1999)}]{1999PASP..111...63F}
{Fitzpatrick}, E.~L. 1999, \pasp, 111, 63, \dodoi{10.1086/316293}

\bibitem[{{Fitzpatrick} \& {Massa}(1986)}]{1986ApJ...307..286F}
{Fitzpatrick}, E.~L., \& {Massa}, D. 1986, \apj, 307, 286,
  \dodoi{10.1086/164415}

\bibitem[{{Fitzpatrick} \& {Massa}(1988)}]{1988ApJ...328..734F}
---. 1988, \apj, 328, 734, \dodoi{10.1086/166332}

\bibitem[{{Fitzpatrick} \& {Massa}(2005)}]{2005AJ....130.1127F}
---. 2005, \aj, 130, 1127, \dodoi{10.1086/431900}

\bibitem[{{Fitzpatrick} \& {Massa}(2007)}]{2007ApJ...663..320F}
---. 2007, \apj, 663, 320, \dodoi{10.1086/518158}

\bibitem[{{Fitzpatrick} \& {Massa}(2009)}]{2009ApJ...699.1209F}
---. 2009, \apj, 699, 1209, \dodoi{10.1088/0004-637X/699/2/1209}

\bibitem[{{Fitzpatrick} {et~al.}(2019){Fitzpatrick}, {Massa}, {Gordon},
  {Bohlin}, \& {Clayton}}]{2019ApJ...886..108F}
{Fitzpatrick}, E.~L., {Massa}, D., {Gordon}, K.~D., {Bohlin}, R., \& {Clayton},
  G.~C. 2019, \apj, 886, 108, \dodoi{10.3847/1538-4357/ab4c3a}

\bibitem[{{Foreman-Mackey} {et~al.}(2013){Foreman-Mackey}, {Hogg}, {Lang}, \&
  {Goodman}}]{2013PASP..125..306F}
{Foreman-Mackey}, D., {Hogg}, D.~W., {Lang}, D., \& {Goodman}, J. 2013, \pasp,
  125, 306, \dodoi{10.1086/670067}

\bibitem[{{Fritz} {et~al.}(2011){Fritz}, {Gillessen}, {Dodds-Eden}, {Lutz},
  {Genzel}, {Raab}, {Ott}, {Pfuhl}, {Eisenhauer}, \&
  {Yusef-Zadeh}}]{2011ApJ...737...73F}
{Fritz}, T.~K., {Gillessen}, S., {Dodds-Eden}, K., {et~al.} 2011, \apj, 737,
  73, \dodoi{10.1088/0004-637X/737/2/73}

\bibitem[{{Gaia Collaboration}(2018)}]{2018yCat.1345....0G}
{Gaia Collaboration}. 2018, VizieR Online Data Catalog, I/345

\bibitem[{{Garrison} \& {Kormendy}(1976)}]{1976PASP...88..865G}
{Garrison}, R.~F., \& {Kormendy}, J. 1976, \pasp, 88, 865,
  \dodoi{10.1086/130037}

\bibitem[{Gordon \& Decleir(2021)}]{measureextinction}
Gordon, K., \& Decleir, M. 2021, {karllark/measure\_extinction: NIR SpeX paper
  release}, v1.2,  Zenodo, \dodoi{10.5281/zenodo.5806655}

\bibitem[{Gordon {et~al.}(2022)Gordon, Larson, McBride, Lim, Sipőcz, \&
  Gaikwad}]{dustextinction}
Gordon, K., Larson, K., McBride, A., {et~al.} 2022, {karllark/dust\_extinction:
  NIRSpectralExtinctionAdded}, v1.1,  Zenodo, \dodoi{10.5281/zenodo.6397654}

\bibitem[{{Gordon} {et~al.}(2009){Gordon}, {Cartledge}, \&
  {Clayton}}]{2009ApJ...705.1320G}
{Gordon}, K.~D., {Cartledge}, S., \& {Clayton}, G.~C. 2009, \apj, 705, 1320,
  \dodoi{10.1088/0004-637X/705/2/1320}

\bibitem[{{Gordon} {et~al.}(2003){Gordon}, {Clayton}, {Misselt}, {Landolt}, \&
  {Wolff}}]{2003ApJ...594..279G}
{Gordon}, K.~D., {Clayton}, G.~C., {Misselt}, K.~A., {Landolt}, A.~U., \&
  {Wolff}, M.~J. 2003, \apj, 594, 279, \dodoi{10.1086/376774}

\bibitem[{{Gordon} {et~al.}(2021){Gordon}, {Misselt}, {Bouwman}, {Clayton},
  {Decleir}, {Hines}, {Pendleton}, {Rieke}, {Smith}, \&
  {Whittet}}]{2021ApJ...916...33G}
{Gordon}, K.~D., {Misselt}, K.~A., {Bouwman}, J., {et~al.} 2021, \apj, 916, 33,
  \dodoi{10.3847/1538-4357/ac00b7}

\bibitem[{{Guetter}(1968)}]{1968PASP...80..197G}
{Guetter}, H.~H. 1968, \pasp, 80, 197, \dodoi{10.1086/128611}

\bibitem[{{Guetter}(1974)}]{1974PASP...86..795G}
---. 1974, \pasp, 86, 795, \dodoi{10.1086/129675}

\bibitem[{{Hanson} {et~al.}(1996){Hanson}, {Conti}, \&
  {Rieke}}]{1996ApJS..107..281H}
{Hanson}, M.~M., {Conti}, P.~S., \& {Rieke}, M.~J. 1996, \apjs, 107, 281,
  \dodoi{10.1086/192366}

\bibitem[{{Hiltner}(1956)}]{1956ApJS....2..389H}
{Hiltner}, W.~A. 1956, \apjs, 2, 389, \dodoi{10.1086/190029}

\bibitem[{{Houk} \& {Swift}(1999)}]{1999MSS...C05....0H}
{Houk}, N., \& {Swift}, C. 1999, Michigan Spectral Survey, 5, 0

\bibitem[{{Indebetouw} {et~al.}(2005){Indebetouw}, {Mathis}, {Babler}, {Meade},
  {Watson}, {Whitney}, {Wolff}, {Wolfire}, {Cohen}, {Bania}, {Benjamin},
  {Clemens}, {Dickey}, {Jackson}, {Kobulnicky}, {Marston}, {Mercer},
  {Stauffer}, {Stolovy}, \& {Churchwell}}]{2005ApJ...619..931I}
{Indebetouw}, R., {Mathis}, J.~S., {Babler}, B.~L., {et~al.} 2005, \apj, 619,
  931, \dodoi{10.1086/426679}

\bibitem[{{Johnson}(1965)}]{1965ApJ...141..923J}
{Johnson}, H.~L. 1965, \apj, 141, 923, \dodoi{10.1086/148186}

\bibitem[{{Johnson} {et~al.}(1966){Johnson}, {Mitchell}, {Iriarte}, \&
  {Wisniewski}}]{1966CoLPL...4...99J}
{Johnson}, H.~L., {Mitchell}, R.~I., {Iriarte}, B., \& {Wisniewski}, W.~Z.
  1966, Communications of the Lunar and Planetary Laboratory, 4, 99

\bibitem[{{Johnson} \& {Morgan}(1955)}]{1955ApJ...122..429J}
{Johnson}, H.~L., \& {Morgan}, W.~W. 1955, \apj, 122, 429,
  \dodoi{10.1086/146103}

\bibitem[{{Jones} {et~al.}(2013){Jones}, {Fanciullo}, {K{\"o}hler},
  {Verstraete}, {Guillet}, {Bocchio}, \& {Ysard}}]{2013A&A...558A..62J}
{Jones}, A.~P., {Fanciullo}, L., {K{\"o}hler}, M., {et~al.} 2013, \aap, 558,
  A62, \dodoi{10.1051/0004-6361/201321686}

\bibitem[{{Jones} \& {Hyland}(1980)}]{1980MNRAS.192..359J}
{Jones}, T.~J., \& {Hyland}, A.~R. 1980, \mnras, 192, 359,
  \dodoi{10.1093/mnras/192.3.359}

\bibitem[{{Knacke} {et~al.}(1982){Knacke}, {McCorkle}, {Puetter}, {Erickson},
  \& {Kraetschmer}}]{1982ApJ...260..141K}
{Knacke}, R.~F., {McCorkle}, S., {Puetter}, R.~C., {Erickson}, E.~F., \&
  {Kraetschmer}, W. 1982, \apj, 260, 141, \dodoi{10.1086/160241}

\bibitem[{{Lanz} \& {Hubeny}(2003)}]{2003ApJS..146..417L}
{Lanz}, T., \& {Hubeny}, I. 2003, \apjs, 146, 417, \dodoi{10.1086/374373}

\bibitem[{{Lesh}(1968)}]{1968ApJS...17..371L}
{Lesh}, J.~R. 1968, \apjs, 17, 371, \dodoi{10.1086/190179}

\bibitem[{{Levato}(1975)}]{1975A&AS...19...91L}
{Levato}, H. 1975, \aaps, 19, 91

\bibitem[{{Lord}(1992)}]{1992nstc.rept.....L}
{Lord}, S.~D. 1992, {A new software tool for computing Earth's atmospheric
  transmission of near- and far-infrared radiation}, NASA Technical Memorandum
  103957

\bibitem[{{Lutz} \& {Lutz}(1977)}]{1977AJ.....82..431L}
{Lutz}, T.~E., \& {Lutz}, J.~H. 1977, \aj, 82, 431, \dodoi{10.1086/112066}

\bibitem[{{Ma{\'\i}z Apell{\'a}niz}(2013)}]{2013hsa7.conf..583M}
{Ma{\'\i}z Apell{\'a}niz}, J. 2013, in Highlights of Spanish Astrophysics VII,
  ed. J.~C. {Guirado}, L.~M. {Lara}, V.~{Quilis}, \& J.~{Gorgas}, 583--589.
\newblock \doarXiv{1209.2560}

\bibitem[{{Ma{\'\i}z Apell{\'a}niz} \& {Barb{\'a}}(2018)}]{2018A&A...613A...9M}
{Ma{\'\i}z Apell{\'a}niz}, J., \& {Barb{\'a}}, R.~H. 2018, \aap, 613, A9,
  \dodoi{10.1051/0004-6361/201732050}

\bibitem[{{Ma{\'\i}z Apell{\'a}niz} {et~al.}(2020){Ma{\'\i}z Apell{\'a}niz},
  {Pantaleoni Gonz{\'a}lez}, {Barb{\'a}}, {Garc{\'\i}a-Lario}, \&
  {Nogueras-Lara}}]{2020MNRAS.496.4951M}
{Ma{\'\i}z Apell{\'a}niz}, J., {Pantaleoni Gonz{\'a}lez}, M., {Barb{\'a}},
  R.~H., {Garc{\'\i}a-Lario}, P., \& {Nogueras-Lara}, F. 2020, \mnras, 496,
  4951, \dodoi{10.1093/mnras/staa1790}

\bibitem[{{Ma{\'\i}z Apell{\'a}niz} \& {Rubio}(2012)}]{2012A&A...541A..54M}
{Ma{\'\i}z Apell{\'a}niz}, J., \& {Rubio}, M. 2012, \aap, 541, A54,
  \dodoi{10.1051/0004-6361/201118712}

\bibitem[{{Martin} \& {Whittet}(1990)}]{1990ApJ...357..113M}
{Martin}, P.~G., \& {Whittet}, D.~C.~B. 1990, \apj, 357, 113,
  \dodoi{10.1086/168896}

\bibitem[{{Massa} {et~al.}(2020){Massa}, {Fitzpatrick}, \&
  {Gordon}}]{2020ApJ...891...67M}
{Massa}, D., {Fitzpatrick}, E.~L., \& {Gordon}, K.~D. 2020, \apj, 891, 67,
  \dodoi{10.3847/1538-4357/ab6f01}

\bibitem[{{Massa} {et~al.}(1983){Massa}, {Savage}, \&
  {Fitzpatrick}}]{1983ApJ...266..662M}
{Massa}, D., {Savage}, B.~D., \& {Fitzpatrick}, E.~L. 1983, \apj, 266, 662,
  \dodoi{10.1086/160813}

\bibitem[{{Mathis} \& {Cardelli}(1992)}]{1992ApJ...398..610M}
{Mathis}, J.~S., \& {Cardelli}, J.~A. 1992, \apj, 398, 610,
  \dodoi{10.1086/171886}

\bibitem[{{Megier} {et~al.}(2009){Megier}, {Strobel}, {Galazutdinov}, \&
  {Kre{\l}owski}}]{2009A&A...507..833M}
{Megier}, A., {Strobel}, A., {Galazutdinov}, G.~A., \& {Kre{\l}owski}, J. 2009,
  \aap, 507, 833, \dodoi{10.1051/0004-6361/20079144}

\bibitem[{{Mendoza}(1967)}]{1967BOTT....4..149M}
{Mendoza}, E.~E. 1967, Boletin de los Observatorios Tonantzintla y Tacubaya, 4,
  149

\bibitem[{{Menzies} {et~al.}(1990){Menzies}, {Marang}, \&
  {Westerhuys}}]{1990SAAOC..14...33M}
{Menzies}, J.~W., {Marang}, F., \& {Westerhuys}, J.~E. 1990, South African
  Astronomical Observatory Circular, 14, 33

\bibitem[{{Morgan} {et~al.}(1955){Morgan}, {Code}, \&
  {Whitford}}]{1955ApJS....2...41M}
{Morgan}, W.~W., {Code}, A.~D., \& {Whitford}, A.~E. 1955, \apjs, 2, 41,
  \dodoi{10.1086/190016}

\bibitem[{{Murakawa} {et~al.}(2000){Murakawa}, {Tamura}, \&
  {Nagata}}]{2000ApJS..128..603M}
{Murakawa}, K., {Tamura}, M., \& {Nagata}, T. 2000, \apjs, 128, 603,
  \dodoi{10.1086/313387}

\bibitem[{{Nishiyama} {et~al.}(2009){Nishiyama}, {Tamura}, {Hatano}, {Kato},
  {Tanab{\'e}}, {Sugitani}, \& {Nagata}}]{2009ApJ...696.1407N}
{Nishiyama}, S., {Tamura}, M., {Hatano}, H., {et~al.} 2009, \apj, 696, 1407,
  \dodoi{10.1088/0004-637X/696/2/1407}

\bibitem[{{Nishiyama} {et~al.}(2006){Nishiyama}, {Nagata}, {Kusakabe},
  {Matsunaga}, {Naoi}, {Kato}, {Nagashima}, {Sugitani}, {Tamura}, {Tanab{\'e}},
  \& {Sato}}]{2006ApJ...638..839N}
{Nishiyama}, S., {Nagata}, T., {Kusakabe}, N., {et~al.} 2006, \apj, 638, 839,
  \dodoi{10.1086/499038}

\bibitem[{{Nogueras-Lara} {et~al.}(2019){Nogueras-Lara}, {Sch{\"o}del},
  {Najarro}, {Gallego-Calvente}, {Gallego-Cano}, {Shahzamanian}, \&
  {Neumayer}}]{2019AA...630L...3N}
{Nogueras-Lara}, F., {Sch{\"o}del}, R., {Najarro}, F., {et~al.} 2019, \aap,
  630, L3, \dodoi{10.1051/0004-6361/201936322}

\bibitem[{{Nogueras-Lara} {et~al.}(2018){Nogueras-Lara}, {Gallego-Calvente},
  {Dong}, {Gallego-Cano}, {Girard}, {Hilker}, {de Zeeuw}, {Feldmeier-Krause},
  {Nishiyama}, {Najarro}, {Neumayer}, \& {Sch{\"o}del}}]{2018AA...610A..83N}
{Nogueras-Lara}, F., {Gallego-Calvente}, A.~T., {Dong}, H., {et~al.} 2018,
  \aap, 610, A83, \dodoi{10.1051/0004-6361/201732002}

\bibitem[{{Pendleton} \& {Allamandola}(2002)}]{2002ApJS..138...75P}
{Pendleton}, Y.~J., \& {Allamandola}, L.~J. 2002, \apjs, 138, 75,
  \dodoi{10.1086/322999}

\bibitem[{{Pendleton} {et~al.}(1994){Pendleton}, {Sandford}, {Allamandola},
  {Tielens}, \& {Sellgren}}]{1994ApJ...437..683P}
{Pendleton}, Y.~J., {Sandford}, S.~A., {Allamandola}, L.~J., {Tielens},
  A.~G.~G.~M., \& {Sellgren}, K. 1994, \apj, 437, 683, \dodoi{10.1086/175031}

\bibitem[{{Potapov} {et~al.}(2021){Potapov}, {Bouwman}, {J{\"a}ger}, \&
  {Henning}}]{2021NatAs...5...78P}
{Potapov}, A., {Bouwman}, J., {J{\"a}ger}, C., \& {Henning}, T. 2021, Nature
  Astronomy, 5, 78, \dodoi{10.1038/s41550-020-01214-x}

\bibitem[{{Rayner} {et~al.}(2003){Rayner}, {Toomey}, {Onaka}, {Denault},
  {Stahlberger}, {Vacca}, {Cushing}, \& {Wang}}]{2003PASP..115..362R}
{Rayner}, J.~T., {Toomey}, D.~W., {Onaka}, P.~M., {et~al.} 2003, \pasp, 115,
  362, \dodoi{10.1086/367745}

\bibitem[{{Rieke} \& {Lebofsky}(1985)}]{1985ApJ...288..618R}
{Rieke}, G.~H., \& {Lebofsky}, M.~J. 1985, \apj, 288, 618,
  \dodoi{10.1086/162827}

\bibitem[{{Rieke} {et~al.}(1989){Rieke}, {Rieke}, \&
  {Paul}}]{1989ApJ...336..752R}
{Rieke}, G.~H., {Rieke}, M.~J., \& {Paul}, A.~E. 1989, \apj, 336, 752,
  \dodoi{10.1086/167047}

\bibitem[{{Rodrigo} \& {Solano}(2020)}]{2020sea..confE.182R}
{Rodrigo}, C., \& {Solano}, E. 2020, in Contributions to the XIV.0 Scientific
  Meeting (virtual) of the Spanish Astronomical Society, 182

\bibitem[{{Rodrigo} {et~al.}(2012){Rodrigo}, {Solano}, \&
  {Bayo}}]{2012ivoa.rept.1015R}
{Rodrigo}, C., {Solano}, E., \& {Bayo}, A. 2012, {SVO Filter Profile Service
  Version 1.0}, IVOA Working Draft 15 October 2012,
  \dodoi{10.5479/ADS/bib/2012ivoa.rept.1015R}

\bibitem[{{Sharpless}(1952)}]{1952ApJ...116..251S}
{Sharpless}, S. 1952, \apj, 116, 251, \dodoi{10.1086/145610}

\bibitem[{{Skrutskie} {et~al.}(2006){Skrutskie}, {Cutri}, {Stiening},
  {Weinberg}, {Schneider}, {Carpenter}, {Beichman}, {Capps}, {Chester},
  {Elias}, {Huchra}, {Liebert}, {Lonsdale}, {Monet}, {Price}, {Seitzer},
  {Jarrett}, {Kirkpatrick}, {Gizis}, {Howard}, {Evans}, {Fowler}, {Fullmer},
  {Hurt}, {Light}, {Kopan}, {Marsh}, {McCallon}, {Tam}, {Van Dyk}, \&
  {Wheelock}}]{2006AJ....131.1163S}
{Skrutskie}, M.~F., {Cutri}, R.~M., {Stiening}, R., {et~al.} 2006, \aj, 131,
  1163, \dodoi{10.1086/498708}

\bibitem[{Skrutskie {et~al.}(2019)Skrutskie, Cutri, Stiening, Weinberg,
  Schneider, Carpenter, Beichman, Capps, Chester, Elias, Huchra, Liebert,
  Lonsdale, Monet, Price, Seitzer, Jarrett, Kirkpatrick, Gizis, Howard, Evans,
  Fowler, Fullmer, Hurt, Light, Kopan, Marsh, McCallon, Tam, Van~Dyk, \&
  Wheelock}]{irsa2mass}
Skrutskie, M.~F., Cutri, R.~M., Stiening, R., {et~al.} 2019, 2MASS All-Sky
  Point Source Catalog,  IPAC, \dodoi{10.26131/IRSA2}

\bibitem[{{Slutskij} {et~al.}(1980){Slutskij}, {Stalbovskij}, \&
  {Shevchenko}}]{1980SvAL....6..397S}
{Slutskij}, V.~E., {Stalbovskij}, O.~I., \& {Shevchenko}, V.~S. 1980, Soviet
  Astronomy Letters, 6, 397

\bibitem[{{Smith}(1987)}]{1987MNRAS.227..943S}
{Smith}, R.~G. 1987, \mnras, 227, 943, \dodoi{10.1093/mnras/227.4.943}

\bibitem[{{Smith} {et~al.}(1993){Smith}, {Sellgren}, \&
  {Brooke}}]{1993MNRAS.263..749S}
{Smith}, R.~G., {Sellgren}, K., \& {Brooke}, T.~Y. 1993, \mnras, 263, 749,
  \dodoi{10.1093/mnras/263.3.749}

\bibitem[{{Smith Neubig} \& {Bruhweiler}(1997)}]{1997AJ....114.1951S}
{Smith Neubig}, M.~M., \& {Bruhweiler}, F.~C. 1997, \aj, 114, 1951,
  \dodoi{10.1086/118617}

\bibitem[{{Sota} {et~al.}(2014){Sota}, {Ma{\'\i}z Apell{\'a}niz}, {Morrell},
  {Barb{\'a}}, {Walborn}, {Gamen}, {Arias}, \& {Alfaro}}]{2014ApJS..211...10S}
{Sota}, A., {Ma{\'\i}z Apell{\'a}niz}, J., {Morrell}, N.~I., {et~al.} 2014,
  \apjs, 211, 10, \dodoi{10.1088/0067-0049/211/1/10}

\bibitem[{{Sota} {et~al.}(2011){Sota}, {Ma{\'\i}z Apell{\'a}niz}, {Walborn},
  {Alfaro}, {Barb{\'a}}, {Morrell}, {Gamen}, \& {Arias}}]{2011ApJS..193...24S}
{Sota}, A., {Ma{\'\i}z Apell{\'a}niz}, J., {Walborn}, N.~R., {et~al.} 2011,
  \apjs, 193, 24, \dodoi{10.1088/0067-0049/193/2/24}

\bibitem[{Stancik \& Brauns(2008)}]{STANCIK200866}
Stancik, A.~L., \& Brauns, E.~B. 2008, Vibrational Spectroscopy, 47, 66,
  \dodoi{https://doi.org/10.1016/j.vibspec.2008.02.009}

\bibitem[{{Stead} \& {Hoare}(2009)}]{2009MNRAS.400..731S}
{Stead}, J.~J., \& {Hoare}, M.~G. 2009, \mnras, 400, 731,
  \dodoi{10.1111/j.1365-2966.2009.15530.x}

\bibitem[{{Stecher}(1965)}]{1965ApJ...142.1683S}
{Stecher}, T.~P. 1965, \apj, 142, 1683, \dodoi{10.1086/148462}

\bibitem[{{Thi} {et~al.}(2006){Thi}, {van Dishoeck}, {Dartois}, {Pontoppidan},
  {Schutte}, {Ehrenfreund}, {D'Hendecourt}, \& {Fraser}}]{2006A&A...449..251T}
{Thi}, W.~F., {van Dishoeck}, E.~F., {Dartois}, E., {et~al.} 2006, \aap, 449,
  251, \dodoi{10.1051/0004-6361:20052931}

\bibitem[{{Vacca} {et~al.}(2003){Vacca}, {Cushing}, \&
  {Rayner}}]{2003PASP..115..389V}
{Vacca}, W.~D., {Cushing}, M.~C., \& {Rayner}, J.~T. 2003, \pasp, 115, 389,
  \dodoi{10.1086/346193}

\bibitem[{{Valencic} {et~al.}(2004){Valencic}, {Clayton}, \&
  {Gordon}}]{2004ApJ...616..912V}
{Valencic}, L.~A., {Clayton}, G.~C., \& {Gordon}, K.~D. 2004, \apj, 616, 912,
  \dodoi{10.1086/424922}

\bibitem[{Virtanen {et~al.}(2020)Virtanen, Gommers, Oliphant, Haberland, Reddy,
  Cournapeau, Burovski, Peterson, Weckesser, Bright, {van der Walt}, Brett,
  Wilson, Millman, Mayorov, Nelson, Jones, Kern, Larson, Carey, Polat, Feng,
  Moore, {VanderPlas}, Laxalde, Perktold, Cimrman, Henriksen, Quintero, Harris,
  Archibald, Ribeiro, Pedregosa, {van Mulbregt}, \& {SciPy 1.0
  Contributors}}]{2020SciPy-NMeth}
Virtanen, P., Gommers, R., Oliphant, T.~E., {et~al.} 2020, Nature Methods, 17,
  261, \dodoi{10.1038/s41592-019-0686-2}

\bibitem[{{Weingartner} \& {Draine}(2001)}]{2001ApJ...548..296W}
{Weingartner}, J.~C., \& {Draine}, B.~T. 2001, \apj, 548, 296,
  \dodoi{10.1086/318651}

\bibitem[{{Wenger} {et~al.}(2000){Wenger}, {Ochsenbein}, {Egret}, {Dubois},
  {Bonnarel}, {Borde}, {Genova}, {Jasniewicz}, {Lalo{\"e}}, {Lesteven}, \&
  {Monier}}]{2000A&AS..143....9W}
{Wenger}, M., {Ochsenbein}, F., {Egret}, D., {et~al.} 2000, \aaps, 143, 9,
  \dodoi{10.1051/aas:2000332}

\bibitem[{{Whittet}(2015)}]{2015ApJ...811..110W}
{Whittet}, D.~C.~B. 2015, \apj, 811, 110, \dodoi{10.1088/0004-637X/811/2/110}

\bibitem[{{Whittet} {et~al.}(1988){Whittet}, {Bode}, {Longmore}, {Adamson},
  {McFadzean}, {Aitken}, \& {Roche}}]{1988MNRAS.233..321W}
{Whittet}, D.~C.~B., {Bode}, M.~F., {Longmore}, A.~J., {et~al.} 1988, \mnras,
  233, 321, \dodoi{10.1093/mnras/233.2.321}

\bibitem[{{Whittet} {et~al.}(1997){Whittet}, {Boogert}, {Gerakines}, {Schutte},
  {Tielens}, {de Graauw}, {Prusti}, {van Dishoeck}, {Wesselius}, \&
  {Wright}}]{1997ApJ...490..729W}
{Whittet}, D.~C.~B., {Boogert}, A.~C.~A., {Gerakines}, P.~A., {et~al.} 1997,
  \apj, 490, 729, \dodoi{10.1086/304914}

\bibitem[{{Wright} {et~al.}(2010){Wright}, {Eisenhardt}, {Mainzer}, {Ressler},
  {Cutri}, {Jarrett}, {Kirkpatrick}, {Padgett}, {McMillan}, {Skrutskie},
  {Stanford}, {Cohen}, {Walker}, {Mather}, {Leisawitz}, {Gautier}, {McLean},
  {Benford}, {Lonsdale}, {Blain}, {Mendez}, {Irace}, {Duval}, {Liu}, {Royer},
  {Heinrichsen}, {Howard}, {Shannon}, {Kendall}, {Walsh}, {Larsen}, {Cardon},
  {Schick}, {Schwalm}, {Abid}, {Fabinsky}, {Naes}, \&
  {Tsai}}]{2010AJ....140.1868W}
{Wright}, E.~L., {Eisenhardt}, P. R.~M., {Mainzer}, A.~K., {et~al.} 2010, \aj,
  140, 1868, \dodoi{10.1088/0004-6256/140/6/1868}

\bibitem[{Wright {et~al.}(2019)Wright, Eisenhardt, Mainzer, Ressler, Cutri,
  Jarrett, Kirkpatrick, Padgett, McMillan, Skrutskie, Stanford, Cohen, Walker,
  Mather, Leisawitz, Gautier, McLean, Benford, Lonsdale, Blain, Mendez, Irace,
  Duval, Liu, Royer, Heinrichsen, Howard, Shannon, Kendall, Walsh, Larsen,
  Cardon, Schick, Schwalm, Abid, Fabinsky, Naes, \& Tsai}]{irsawise}
Wright, E.~L., Eisenhardt, P. R.~M., Mainzer, A.~K., {et~al.} 2019, AllWISE
  Source Catalog,  IPAC, \dodoi{10.26131/IRSA1}

\bibitem[{{Zubko} {et~al.}(2004){Zubko}, {Dwek}, \&
  {Arendt}}]{2004ApJS..152..211Z}
{Zubko}, V., {Dwek}, E., \& {Arendt}, R.~G. 2004, \apjs, 152, 211,
  \dodoi{10.1086/382351}

\end{thebibliography}
\bibliographystyle{aasjournal}

\end{document}